\documentclass[fleqn,usenatbib]{mnras}

\usepackage{newtxtext,newtxmath}

\usepackage[T1]{fontenc}
\usepackage{ae,aecompl}

\usepackage{graphicx}	
\usepackage{amsmath}	
\usepackage{amssymb}	
\usepackage{lscape}
\usepackage{natbib}
\usepackage{keyval}
\usepackage{lastpage}
\usepackage{array}
\usepackage{dcolumn}
\usepackage{comment}
\usepackage{pifont}
\usepackage{natbib,times}
\usepackage[T1]{fontenc}
\usepackage{aecompl}

\newcommand{\mstar}{\hbox{${\mathrm M}_\ast$}}
\newcommand{\msun}{\hbox{$\rm{M_{_\odot}}$}}
\newcommand{\zsun}{\hbox{$\rm{Z_{_\odot}}$}}

\def\arcsec{\hbox{$^{\prime\prime}$}}

\def \be {\begin{equation}}
\def \ee {\end{equation}}

\def\gsim{\mathrel{\lower0.6ex\hbox{$\buildrel {\textstyle >}
 \over {\scriptstyle \sim}$}}}
\def\lsim{\mathrel{\lower0.6ex\hbox{$\buildrel {\textstyle <}
 \over {\scriptstyle \sim}$}}}

\def \halpha {H$\alpha$}

\def \hb {H$\beta$}

\title[SDSS-IV MaNGA: Spatially resolved star-formation histories]{SDSS-IV MaNGA: Spatially resolved star-formation histories and the connection to galaxy physical properties}

\author[K. Rowlands et al.]{
K. Rowlands$^{1}$\thanks{E-mail:katerowlands.astro@gmail.com}, T. Heckman$^1$, V. Wild$^2$, N. L. Zakamska$^1$, V. Rodriguez-Gomez$^1$,
\newauthor J. Barrera-Ballesteros$^1$, J. Lotz$^{3,1}$, D. Thilker$^1$,
B. H. Andrews$^4$, M. Boquien$^5$,
\newauthor J. Brinkmann$^6$, J. R. Brownstein$^7$, H-C. Hwang$^1$, R. Smethurst$^8$ \\
$^1$Department of Physics \& Astronomy, Johns Hopkins University, Bloomberg centre, 3400 N. Charles St., Baltimore, MD 21218, USA \\
$^2$School of Physics \& Astronomy, University of St Andrews, North Haugh, St Andrews, Fife, KY16 9SS, UK \\
$^3$ Space Telescope Science Institute, 3700 San Martin Dr., Baltimore, MD 21218, USA \\
$^4$PITT PACC, Department of Physics and Astronomy, University of Pittsburgh, Pittsburgh, PA 15260, USA \\
$^5$Centro de Astronomía (CITEVA), Universidad de Antofagasta, Avenida Angamos 601, Antofagasta, Chile \\
$^6$Apache Point Observatory, P.O. Box 59, Sunspot, NM 88349 \\
$^7$Department of Physics and Astronomy, University of Utah, 115 S. 1400 E., Salt Lake City, UT 84112, USA \\
$^8$Centre for Astronomy \& Particle Theory, University Park, Nottingham, NG7 2RD, UK \\
}

\date{Accepted XXX. Received YYY; in original form ZZZ}

\pubyear{2018}

\begin{document}
\label{firstpage}
\pagerange{\pageref{firstpage}--\pageref{lastpage}}
\maketitle

\begin{abstract}
A key task of observational extragalactic astronomy is to determine where -- within galaxies of diverse masses and morphologies -- stellar mass growth occurs, how it depends on galaxy properties and what processes regulate star formation. Using spectroscopic indices derived from the stellar continuum at $\sim 4000$\AA, we determine the spatially resolved star-formation histories of 980000 spaxels in 2404 galaxies in the SDSS-IV MaNGA IFU survey. We examine the spatial distribution of star-forming, quiescent, green valley, starburst and post-starburst spaxels as a function of stellar mass and morphology to see where and in what types of galaxy star formation is occurring. The spatial distribution of star-formation is dependent primarily on stellar mass, with a noticeable change in the distribution at \mstar$>10^{10}$\msun. Galaxies above this mass have an increasing fraction of regions that are forming stars with increasing radius, whereas lower mass galaxies have a constant fraction of star forming regions with radius. Our findings support a picture of inside-out growth and quenching at high masses.
We find that morphology (measured via concentration) correlates with the fraction of star-forming spaxels, but not with their radial distribution. We find (post-)starburst regions are more common outside of the galaxy centre, are preferentially found in asymmetric galaxies, and have lower gas-phase metallicity than other regions, consistent with interactions triggering starbursts and driving low metallicity gas into regions at $<1.5R_e$.
\end{abstract}

\begin{keywords}
galaxies: evolution -- galaxies: interactions -- galaxies: abundances -- galaxies: ISM -- galaxies: starburst
\end{keywords}



\section{Introduction}

How and where galaxies build up their stellar mass over time is key to understanding galaxy evolution.
Galaxies can grow in stellar mass through the conversion of gas into stars (in-situ), and by accretion of stars via galaxy mergers and tidal stripping of smaller satellite galaxies (ex-situ). Simulations suggest that the majority of galaxies in the Universe (with stellar masses below a few times $10^{11}$\msun) grow primarily through in-situ star formation \citep{Rodriguez-Gomez16}. Regions in such galaxies will only continue to grow where gas is available and  conditions are favourable for star formation.

In the hierarchical model of galaxy evolution, galaxies are thought to build their mass from the inside-out (termed \emph{inside-out growth}) \citep[e.g.][]{White_Frenk91, Kauffmann93, Mo98, Roskar08} because gas cools and forms stars first in the centres of growing halos.
In a different but related process, massive galaxies ($\mstar \gtrsim 10^{10}$\msun) are also thought to undergo \emph{inside-out quenching}, with star-formation stopping first in the centre (due to numerous possible processes) and subsequently at larger radius.

This theoretical picture is broadly supported by observations for high mass galaxies
\citep{GonzalezDelgado15, Perez13, Pan15, GonzalezDelgado17}.
Using specific star formation rate (SSFR) profiles of galaxies in the local Universe derived from UV-optical colours, \citet{Munoz_Mateos07} found that galaxies are forming stars at higher relative rates in their outer regions compared to their centres.
Many studies have found negative age or colour gradients in local disc galaxies \citep{Wang11, Perez13, Sanchez-Blazquez14, Goddard17}, suggesting mass built up first in the centre, which subsequently stopped forming stars, whilst star-formation continued at larger radius.
Early-type galaxies tend to exhibit flatter age gradients \citep{GonzalezDelgado15, Goddard17} but are also consistent with growing from the inside-out. Using spatially resolved broad and narrow band/grism derived colours and measurements of the stellar mass and SFR densities, studies have found evidence for inside-out galaxy growth out to $z\sim2$ in star-forming galaxies \citep{Bezanson09, Wuyts12, Nelson12, Nelson13, vandeSande13, Patel13, Tacchella15, Nelson16, Tacchella16a}.

However, current observations suggest that the radial growth of galaxies is mass dependent: in local lower mass disc galaxies star formation is found to proceed from the outside-in or equally at all radii \citep{Perez13, Gallart08, Zhang12, Goddard17a, Sanchez-Blazquez14, GonzalezDelgado15, GonzalezDelgado17, Ibarra-Medel16}. There may also be other hidden factors: \citet{Wang17a} found that some high mass spiral galaxies grow outside-in, and they have smaller sizes, higher stellar surface mass density, higher SFR and higher oxygen abundance than normal star-forming galaxies of a similar stellar mass.

Previous spectroscopic studies of the stellar populations of galaxies were mostly limited to spatially unresolved studies using single fibre spectroscopy \citep[e.g.][]{Kauffmann03a, Brinchmann04, Gallazzi05}. Fibre spectroscopy only samples the inner parts of the galaxy and is prone to aperture bias depending on the redshift range of the sample \citep{Kewley05, Pracy12, Brough13, Richards16}. Whilst broadband photometry can be used for resolved studies \citep[e.g.][]{BelldeJong00, Munoz_Mateos07, Wang11, Fang12}, detailed studies of the stellar populations is limited due to the larger uncertainty on parameters such as stellar population age. Initial integral-field-unit (IFU) surveys focused on small numbers of specifically selected galaxies (\citealp[e.g. SAURON,][]{Bacon01, Davies01}; ATLAS$^{3 \mathrm{D}}$, \citealp{Cappellari11}; DiskMass, \citealp{Bershady10}). Long slit spectroscopic studies of galaxies were undertaken for more representative samples of galaxies by \citet{Moran12} and \citet{Huang13}, however samples were limited to relatively small numbers ($\sim100$) with a limited mass range ($\mstar>10^{10}\msun$).

For the first time, multi-object IFU observations from the Calar Alto Legacy Integral Field Area Survey (CALIFA), Sydney-AAO Multi-object Integral field spectrograph (SAMI) and Mapping Nearby Galaxies at APO (MaNGA) surveys are allowing the study of the spatially resolved stellar populations and ionised gas in large, unbiased samples of galaxies spanning a wide mass range ($10^{9}<\mstar<10^{12}\msun$). While there is consensus in some aspects on where stellar mass growth occurs, there still remain disagreements in detail. Most studies determining stellar mass growth now rely on full-spectrum fitting with stellar population synthesis models to obtain the star formation histories of galaxy regions. Using CALIFA data, \citet{GonzalezDelgado15} measured the spatially resolved mass and light-weighted ages for 300 galaxies spanning a range of masses and Hubble types. They found negative light-weighted age gradients (inside-out formation) on average in galaxies, and that these were steepest in Sb-Sbc galaxies. Using MaNGA data \citet{Ibarra-Medel16} found that star-forming/late-type galaxies have on average a stronger inside-out formation than quiescent/early-type galaxies, independent of mass; similar results were found in simulated Milky-Way like galaxies \citep{Avila-Reese18}. However, they found that low-mass galaxies show diverse radial mass growth histories and have stochastic SFHs, with no clear evidence for outside-in formation as seen in some previous studies \citep{Wang18}. \citet{Goddard17a} used the {\sc{FIREFLY}} full spectrum fitting package on MaNGA galaxies to show that late-type galaxies have negative light-weighted age gradients, again suggesting `inside-out' formation for discs. On the other hand, they found positive age gradients for early-type galaxies, indicating outside-in progression of star formation. Again using MaNGA data, \citet{Wang18} found evidence for strong gradients in spectral indices, indicating inside-out growth in massive galaxies ($\mstar > 10^{10}$\msun), but flat gradients in lower mass galaxies. Finally, \citet{Spindler18} measured the SSFR profiles of intermediate and high mass MaNGA galaxies and found that galaxies with high velocity dispersion or S\'ersic index (i.e. significant bulges) had centrally suppressed star-formation. Similarly, using SAMI data, \citet{Medling18} found that early-type galaxies below the SFR-\mstar relation had lower SSFR and star formation rate surface densities ($\Sigma$SFR) in their centres, indicating inside-out quenching. These latter studies used emission lines as instantaneous star formation measures instead of inferring the star-formation history from stellar populations.

Many studies which derive the mass growth history rely on full spectrum fitting to determine stellar population age and metallicity. These properties are model dependant, and require relatively high signal-to-noise ratio (SNR) spectra and good spectro-photometric calibration. We take a less model-dependant approach by defining spectral indices using a Principal Component Analysis performed for the first time on spatially resolved spectra, which is suitable for use at lower SNR than traditional spectral indices \citep{Wild07}. This allows us to easily visualise the spatially resolved star-formation histories of galaxies and the connection to global and spatially resolved galaxy properties for the largest sample of galaxies with IFU observations to date. We are able to see where and in what types of galaxy (e.g. mass, morphology) stellar mass is currently being built via star formation. In particular, we are able for the first time to identify starburst and post-starburst regions indicating rapid growth and recent quenching, and potentially  different conditions for star formation. We adopt a cosmology with $\Omega_m=0.30,\,\Omega_{\Lambda}=0.70$ and $H_o=70\,\rm{km\,s^{-1}\,Mpc^{-1}}$.

\section{Data and methods}\label{sec:data}

Mapping Nearby Galaxies at APO (MaNGA, \citealp{Bundy15}) is an Integral-Field-Unit (IFU) survey of 10000 galaxies undertaken as part of SDSS-IV \citet{Blanton17}. Observations are undertaken using the BOSS spectrograph \citep{Smee13, Drory15} on the 2.5 meter SDSS telescope at Apache Point Observatory (APO) \citet{Gunn06}. IFUs are comprised of 19 fibres (12\arcsec\, diameter on sky) to 127 fibres (32\arcsec) hexagonal fibre bundles, with each fibre having a diameter of 2\arcsec. Standard stars for flux calibration are observed using 12 mini-bundles of 7 fibres. An additional 96 fibres per plate are used to obtain sky spectra. A three-point dithering pattern is used to ensure full coverage of the field of view \citep{Law16}, with a resulting point spread function (PSF) of $\sim 2.5$\arcsec\, FWHM. The spectra have a wavelength range of 3600--10300\AA, with $R\sim1100-2200$ \citep{Smee13}. The MaNGA data are flux and wavelength calibrated, and sky subtracted using the Data Reduction Pipeline (DRP) version v2\_0\_1. This provides us with datacubes with a spaxel size of 0.5\arcsec and a SNR of 4--8\AA$^{-1}$ in the outer parts of each primary sample galaxy.

The MaNGA sample has a redshift range of $0.01<z<0.18$, and was selected to have a flat distribution in SDSS $i$-band magnitude and therefore be relatively unbiased in stellar mass. The Primary sample comprises 2/3 of the full MaNGA sample and observes galaxies out to $1.5 R_e$. The other 1/3 of the sample is denoted as the Secondary sample and observes galaxies out to $2.5 R_e$. For further details of the survey design and observing strategy see \citet{Law15, Yan15, Wake17}.

\subsection{Sample selection}
\label{sec:sample}
In this work we use the MaNGA collaboration internal release MPL-5 dataset which comprises 2779 galaxies. We exclude 37 galaxies which had bad quality flags (DRP3QUAL$>300$), which reduces our sample to 2742 galaxies. We use both the MaNGA Primary and Secondary samples with Petrosian effective area greater than 4 times the area of the $2.5$\arcsec{} PSF to exclude small galaxies with few spaxels, and one galaxy with an incorrect Petrosian radius. Furthermore, the small galaxies which we exclude would have highly correlated spaxels which reduces our ability to reliably look for radial trends. Furthermore, we make a cut on axial ratio $b/a>0.35$ to exclude edge-on galaxies which may have unreliable spectral classifications due to their high dust column density. Our final sample comprises 2404 galaxies.

\subsection{Physical quantities}
Physical properties such as redshift ($z$), Galactic extinction, total stellar mass (\mstar, using a \citet{Chabrier03} IMF) and broadband $FUV$, $NUV$, $u$, $g$, $r$, $i$ and $z$ magnitudes are taken from the NASA Sloan ATLAS (NSA) {\sc v1\_0\_1} \citep{Blanton11}. The NSA also contains quantities such as photometric position angle (PA), axial ratio ($b/a$), and effective radius ($R_\mathrm{eff}$) which are measured on the SDSS $r$-band images.

The stellar mass density maps are from the Pipe3D pipeline \citep{Sanchez16a, Sanchez16b}. Firstly, the MaNGA cubes are Voronoi binned to a target SNR of 50. The code fits a reduced stellar template set to derive the kinematics. Pipe3D then fits a linear combination of dust attenuated SSP templates and emission line models to each spaxel. Pipe3D uses the stellar library of 156 templates described in \citet{CidFernandes13}, covering 39 stellar ages (1\,Myr to 14.2\,Gyr), and four metallicities ($Z/\zsun = 0.2$, 0.4, 1 and 1.5). Templates are from a combination of the synthetic stellar spectra from the GRANADA library \citep{Martins05} and the MILES SSP library \citep{Sanchez-Blazquez06, Vazdekis10, Falcon-Barroso11}, using a Salpeter initial mass function (IMF). The stellar mass density of each spaxel is derived by accounting for the mass-to-light ratio of each individual template within the library, the stellar mass loss, the weight in light of each component derived from the fitting and the internal dust-corrected surface brightness at each particular look-back time.
We have stellar mass density maps for 2501/2742 galaxies. 241 cubes were excluded which had issues with the data (e.g. low S/N, masked areas, problems with the spectrophotometric calibration, or unsatisfactory results from Pipe3D.) Furthermore, we mask spaxels which have S/N$<20$.
We correct the stellar mass density for the effects of inclination by multiplying by the axial radio ($b/a$) from the NSA catalogue following \citet{Barrera-Ballesteros16}.

We derive the gas mass density using the empirical relation between the gas mass density and the attenuation \citet[][Barrera-Ballesteros et al. in prep]{Barrera-Ballesteros18}

\begin{equation}
\Sigma_\mathrm{gas} = 30 \left(\frac{A_{\mathrm{v}}}{\mathrm{mag}}\right) (\msun \mathrm{pc}^{-2}).
\end{equation}

This relationship is calibrated using the observed gas mass density from spatially resolved observations of CO and the v-band attenuation (A$_\mathrm{v}$) for galaxies in the CALIFA EDGE survey \citep{Bolatto17}. A$_\mathrm{v}$ is measured from the Balmer decrement following \citet{Catalan-Torrecilla15} using the \halpha/\hb\, flux ratio, assuming a canonical value of 2.86 and the \citet{Cardelli89} extinction curve with $R_{\mathrm{v}} = 3.1$. We correct the gas mass density for the effects of inclination by multiplying by the axial radio ($b/a$).

The gas phase metallicity is derived using the O2N2 diagnostic using the \textsc{pyqz} code \citep{Dopita13, pyqz}. We assume that the HII regions can be modelled as spherically symmetric, although this assumption does not affect our results. We take care to only use spaxels which are classified as star-forming in the [NII]/\halpha\, BPT diagram (i.e. are below the \citealt{Kewley01} demarcation line), and which have \halpha\, EW$>6$\AA\, \citep{CidFernandes11, Barrera-Ballesteros18}. Gas mass density and metallicity are measured on the unbinned maps. In all quantities which use emission line measurements (i.e. Balmer decrement, metallicity), we only use spaxels which have emission lines with S/N$>3$ to ensure reliable measurements. Emission lines are dust corrected using the Balmer decrement before input into \textsc{pyqz}.
When counting binned spaxels, we make sure that the central position in each Voronoi bin is associated with the unbinned spaxel in the unbinned maps, so that Voronoi binned spaxels are not doubled counted.

\subsection{Morphology measurements}
In our analysis we utilise non-parametric morphology measurements of concentration ($C$, \citealt{Abraham94, Abraham96}) and asymmetry ($A$, \citealt{Schade95, Abraham96, Wu99, Bershady00, Conselice00}), measured on the Panoramic Survey Telescope and Rapid Response System (Pan-STARRS) i-band Data Release 1 images \citep{Chambers17, Flewelling16}. There are Pan-STARRS images available for 2436 galaxies in MPL-5.
The Pan-STARRS i-band images have a median point source $5\sigma$ depth of 23.1 (AB mag) and have a median PSF of 1.11\arcsec. These images are deeper than the SDSS images and have a smaller PSF, which is why we use them for our morphology measurements. The morphology measurements were performed with the \texttt{statmorph} code (Rodriguez-Gomez et al., in prep) \footnote{https://github.com/vrodgom/statmorph}.
The concentration ($C$) measures the relative amount of flux in a galaxy's central region relative to the outer regions.

\begin{equation}
C = 5 \mathrm{log} \left(\frac{r_{80}}{r_{20}}\right)
\end{equation}

\noindent where $r_{80}$ and $r_{20}$ are the radii containing 80\% and 20\% of the flux, respectively.

The asymmetry parameter ($A$) measures the fraction of light in non-symmetric components.

\begin{equation}
A = \frac{\sum_{i,j} |I(i,j) - I_{180}(i,j)|}{\sum_{i,j} I(i,j)} - A_{B_{180}}
\end{equation}

\noindent where $I$ is the original image and $I_{180}$ is the image rotated by 180 degrees about a central pixel (chosen to minimise $A$). The sum is computed over all pixels within $1.5 R_e$. The subtraction of the asymmetry of the background ($A_{B_{180}}$) accounts for the effect of noise on $A$. The background is measured in a 32x32 pixel region where no structure is detected. Nearby stars are masked before performing the morphology measurements.

We only include reliable morphology measurements for 2047 galaxies where the S/N per pixel is $>2.5$ in the i-band image, the half-light radius is $> 0.5$ times the FHWM of the PSF (1.11\arcsec) and no problems occurred during the morphology measurement.

\subsection{Spectroscopic Sample Classification}
\label{sec:PCA}

In optical spectra, the signatures of stars of different ages can be used to obtain information about the recent star-formation history (SFH). To define our sample we make use of two particular features of optical spectra: the 4000\AA\ break strength and Balmer absorption line strength. Following the method outlined in \citet{Wild07}, we define two spectral indices which are based on a Principal Component Analysis (PCA) of the 3750--4150\AA\ region of the spectra. PC1 is related to the strength of the 4000\AA\ break (equivalent to the $D_{n}4000$ index), and PC2 is the excess Balmer absorption (of all Balmer lines simultaneously) over tht expected for the 4000\AA\ break strength. The eigenbasis that defines the principal components is taken from \citet{Wild07}, and was built using a set of \citet{BC03} model spectra. The set of models assumes exponentially declining star-formation histories (SFHs) with additional superimposed random bursts. The model spectra cover a wide range of age, metallicity and SFH.

To calculate the principal component amplitudes for each spectrum, we correct each MaNGA spectrum for Galactic extinction using the \citet{Cardelli89} extinction law, mask sky lines, shift to rest-frame wavelengths (including removal of small wavelength shifts due to galactic rotation) and interpolate the spectra onto a common wavelength grid. We only use spaxels where the uncertainty on the stellar velocity is $<500$\,km\,s$^{-1}$. We then project each spectrum onto the eigenbasis using the `gappy-PCA' procedure of \citet{Connolly_Szalay99}: pixels are weighted by their errors during the projection, and gaps in the spectra due to bad pixels are given zero weight. The normalisation of the spectra is also free to vary in the projection using the method introduced by \citet{Wild07}.

\begin{figure}
	\includegraphics[width=\columnwidth]{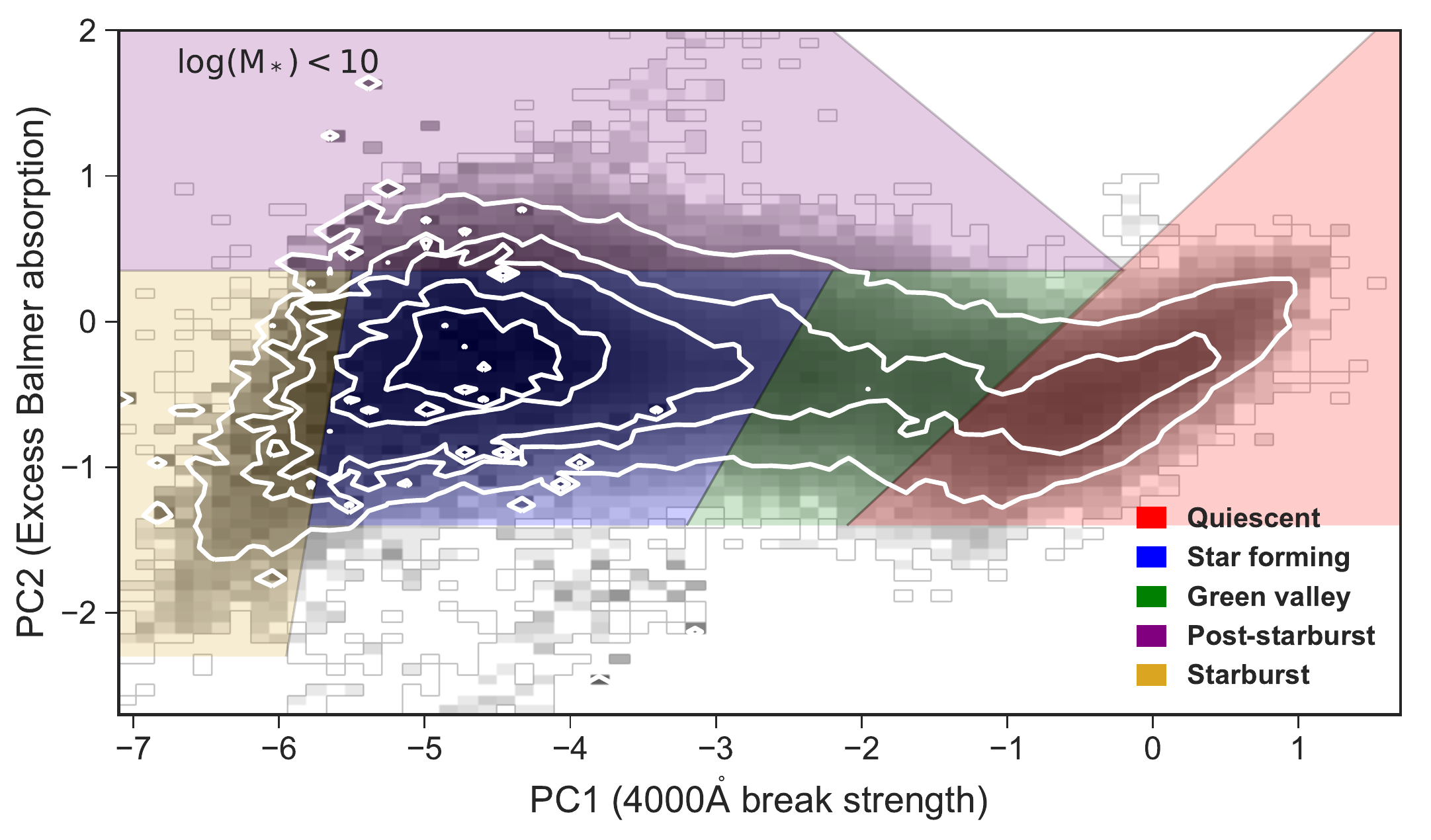}
    \includegraphics[width=\columnwidth]{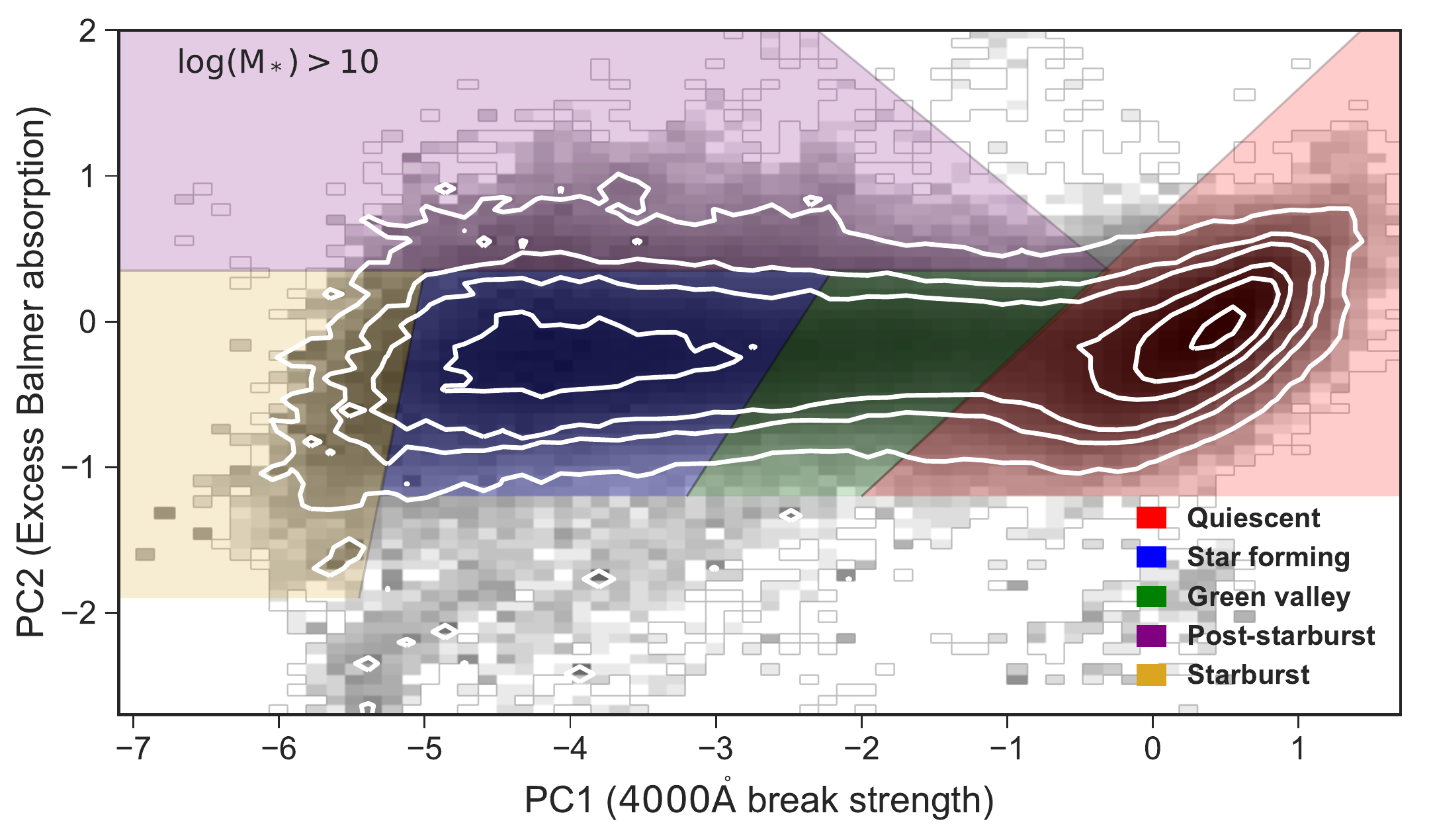}
    \caption{The distribution of the 4000\AA\ break strength (PC1) and excess Balmer absorption (PC2) as measured by a principal component analysis of the 4000\AA\ spectral region of the MaNGA spaxels. The grey-scale indicates the logarithmic number of spaxels. The white contours are 1\%, 5\%, 10\%, 30\%, 50\% and 90\% of the maximum number of spaxels in each mass bin. The coloured regions indicate the classification boundaries: quiescent (red), star forming (blue), green valley (green), starburst (yellow) and post-starburst (purple); these are discussed in detail in Section~\ref{sec:PCA}. The upper and lowers plots show the spaxels in low (log(\mstar)$<10$) and high mass (log(\mstar)$>10$) galaxies, respectively.}
    \label{fig:PC12}
\end{figure}

In Figure~\ref{fig:PC12} we show the distribution of the two spectral indices (PC1, PC2) for spaxels in the MaNGA cubes. A bimodal distribution of the spaxels is clearly evident in Figure~\ref{fig:PC12}, reminiscent of what is observed in integrated light observations. Regions which show no evidence of recent or current star formation lie on the right of Figure~\ref{fig:PC12} (high PC1), as they have a strong 4000\AA\ break. Regions which are forming stars lie in the centre and left of Figure~\ref{fig:PC12} (low PC1). These spectra have younger mean stellar ages and therefore weaker 4000\AA\ breaks. Regions in the more sparsely populated region between the star-forming and quiescent spaxels are defined as green valley (akin to that of the green valley in optical colour-magnitude diagrams).

A small number of spaxels are undergoing a ``starburst'', i.e. there has been a sharp increase in the star formation rate over a short timescale ($\sim10^7$ years).  These spectra are identified by their unusually weak Balmer absorption lines (low PC2), strong UV--blue continua, and weak 4000\AA\ breaks (low PC1) i.e. spectra dominated by light from O/B stars. These spaxels lie in the lower left of Fig. \ref{fig:PC12}. As the starburst ages to a few $10^8$yrs, the Balmer absorption lines increase in strength as the galaxy passes into the post-starburst phase \citep{DresslerGunn83, CouchSharples87}, i.e. A/F star light dominates the spectrum for up to 1\,Gyr following a starburst. These spaxels with stronger Balmer absorption lines lie to the top of Figure~\ref{fig:PC12}. Through comparison with population synthesis models, using simple toy model star formation histories or more complex histories derived from
simulations, \citet{Wild07, Wild09} showed that the shape of the left
hand side of the distribution in Fig. \ref{fig:PC12} describes the
evolutionary track of a starburst, with time since the
starburst increasing from bottom to top, and burst strength increasing
from right to left. The spaxels lying at the outermost edge of the distribution have undergone the strongest recent bursts of star formation in the entire sample. At these low-redshifts, these starbursts are not strong; Bayesian fits to spectral synthesis models imply typical burst mass fractions (i.e. fraction of stellar mass formed in the burst) of $\sim$10\% \citep{Wild10}. The models show that if spaxels that had undergone stronger bursts existed, they would lie to the left of the distribution at intermediate starburst ages, where no spaxels are observed.

\begin{figure*}
	\includegraphics[width=0.92\textwidth]{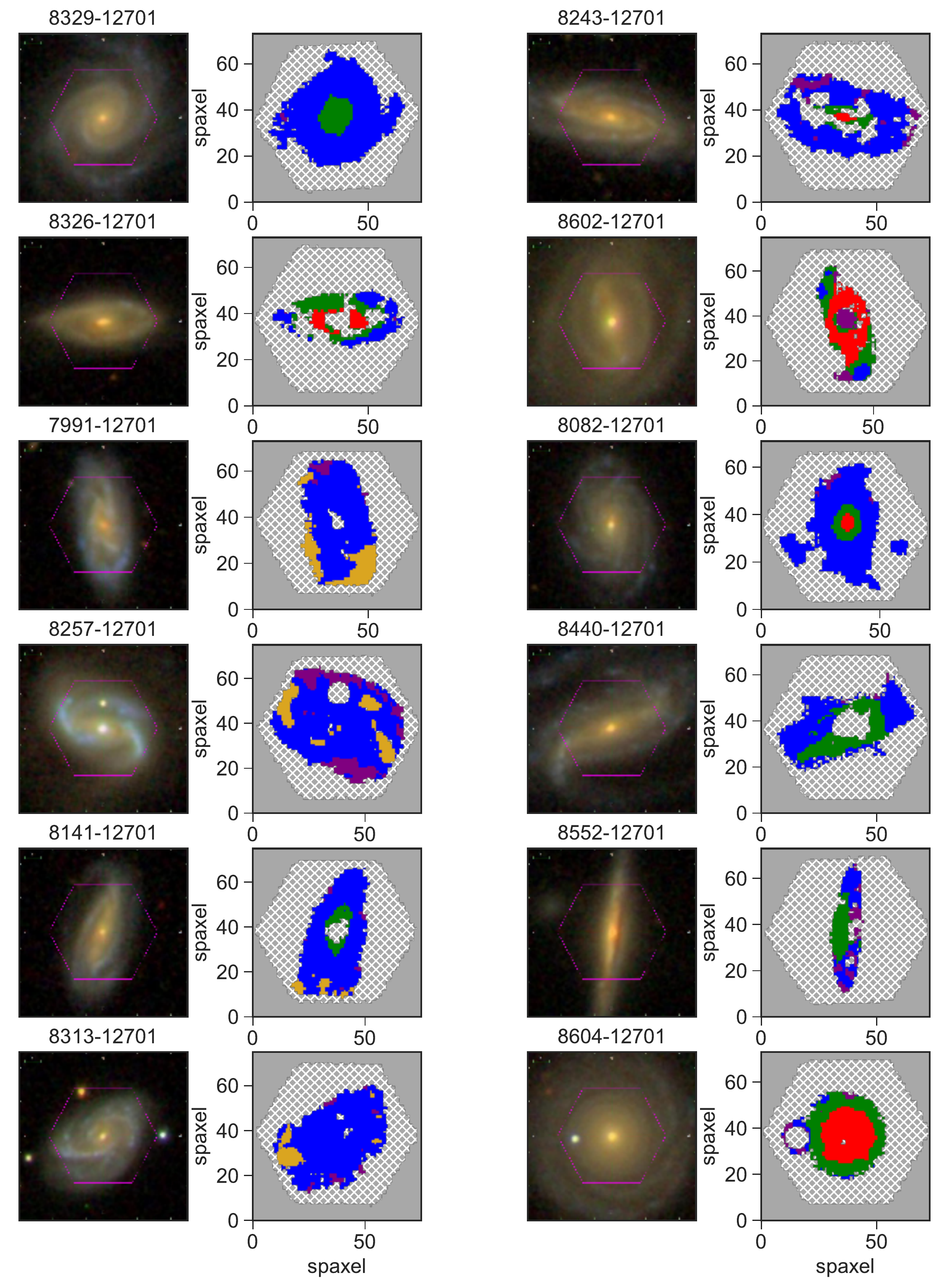}
    \caption{SDSS $gri$ images and maps of the PCA classes of spaxels in example MaNGA galaxies. The IFU footprint is shown as the magenta hexagon. The classes are: quiescent (red), star forming (blue), green valley (green), starburst (yellow) and post-starburst (purple). The grey region indicates no coverage, and the hatched region indicates where spaxels are masked due to low S/N, non-classification due to non-physical values of (PC1, PC2), masking in the DAP cubes e.g. due to stellar contamination, or because of a high Balmer decrement (H$\alpha$/H$\beta$ $>5.16$). Galaxy 8082-12701 is a good example of the importance of MaNGA in overcoming aperture bias present in single fibre measurements: using fibre spectroscopy it would have been classified as quiescent but the disc is dominated by star-forming spaxels.}
\label{fig:Example_PCA}
\end{figure*}

In Figure~\ref{fig:PC12} we divide our sample into five spectral classes based on their values of PC1 and PC2. These boundaries are largely arbitrary, segregating spaxels dominated by broadly similar recent SFH. We move the classification boundaries depending on the stellar mass of the galaxy, because higher mass galaxies have older stellar populations causing them to shift right in PC1. This predominantly affects the starburst boundary, which should be interpreted as selecting the highest SSFR spaxels for galaxies of a given stellar mass. As we are interested in where these spaxels lie, rather than their absolute fraction, this does not affect our analysis. 
Points below the star-forming region are spaxels dominated by broadline active galactic nuclei (AGN) light, or have unusual spectral shapes due to calibration or data reduction issues, examples are shown in Appendix~\ref{sec:Reject_spectra}. These are poorly described by the PCA and are excluded from our analysis. Due to our conservative cuts at low values of PC2 to exclude broadline AGN, a small number of starburst spaxels are removed. We have examined the spaxels near broadline AGN regions by eye in order to check that the contamination by non-stellar light is low.
Our PCA classification of all spaxels allows us to make maps of the recent SFH in each MaNGA galaxy, as shown in Figure~\ref{fig:Example_PCA}.

\begin{figure}
	\includegraphics[width=0.49\textwidth]{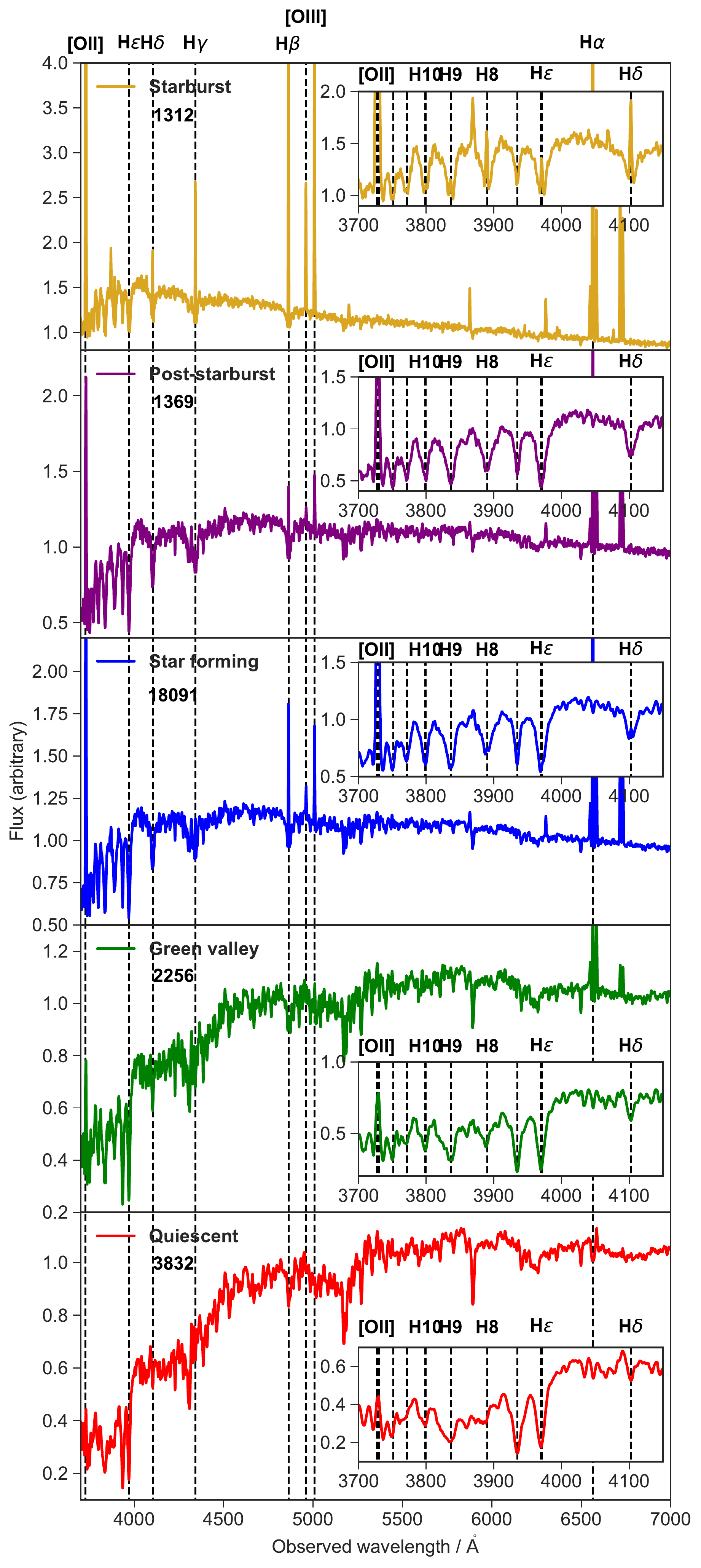}
    \caption{Stacked (median) spectra in each spectral class for 50 representative galaxies. The number of spectra in each stack is shown in each panel. Dotted vertical lines indicate the rest-frame vacuum wavelengths of emission and absorption lines labelled at the top of each panel. The inset plot shows a zoom in on the 4000\AA\, break region which is used for the PCA.}
\label{fig:stacked_spectra}
\end{figure}

Stacking spectra in each class shows that on average the spaxels show the expected characteristic features, see Figure~\ref{fig:stacked_spectra}. Star-forming spaxels show strong emission lines, a weak 4000\AA\ break, and blue continua. The stacked quiescent spaxels show strong 4000\AA\ breaks and weak emission lines. Green valley spaxels show spectral shapes intermediate between those of star-forming and quiescent spaxels with moderately strong 4000\AA\ breaks and weak emission lines. These spectroscopic green valley spaxels do not show characteristic deep Balmer absorption lines, which indicates a slower transition time from star-forming to quiescent, compared to post-starburst spaxels.
Post-starburst spaxels have strong Balmer absorption lines and moderately strong 4000\AA\ breaks. For a spaxel to be classified as post-starburst, the star-formation must have shut-down rapidly (i.e. an order of magnitude drop in SFR over a few hundred Myr \citealt{Wild10, Rowlands15}) within the last 1\,Gyr. 
Note that our spaxel selection method makes no cuts on emission line strength, as is often done in the selection of post-starbursts \citep{Goto05, Goto08}. It is important not to exclude spaxels with emission lines, as narrow line AGN are common in post-starburst samples \citep{Wild07, Yan06, Yan09}, and shocks can excite emission lines in post-starbursts \citep{Alatalo16a}. By making no cuts on emission line strengths our post-starburst class also includes spaxels with some residual star formation, which has not been completely halted following the starburst. This allows us to identify post-starburst galaxies that had their burst more recently or have switched off their star formation more slowly.

The wavelength range for the PCA was chosen following extensive testing on models to be maximally sensitive to recent star formation history, but minimally sensitive to dust. While a longer wavelength range does identify the same properties, the indices are less easy to interpret in terms of physical properties \citep[see e.g.][]{Chen12}.  Emission lines can obviously also be used to split the sample into star-forming and non-starforming spaxels. Spaxels which are classified as star-forming are those below the \citet{Kewley01} demarcation line in [NII]/\halpha\, BPT diagram and which have \halpha\, EW$>6$\AA. Using this method we cannot assign starburst, post-starburst and green valley classifications as emission lines only provide an instantaneous picture. We repeat the analysis in Section 3 using star-forming and non-star-forming spaxels and find qualitatively similar results.

The spectral indices (PC1 and PC2, described in the following section) are calculated on the Voronoi binned (VOR10) maps which are binned to achieve a target SNR of 10 in the $r$-band continuum. Further details of the Voronoi binning, stellar continuum and emission line fitting with the MaNGA Data Analysis Pipeline (DAP) version 2.0.2 are in Westfall et al. (in prep). Note that not all binned spaxels (particularly those in the outer part of a galaxy) reach the target SNR. To ensure reliable  spectral index measurements, we exclude spaxels which have median SNR$<3$ in the 4000\AA\ break region or uncertainty on PC2 $>0.2$ (the 90th percentile of the PC2 error distribution).
We exclude 9\% of classified spaxels which are excessively dusty with Balmer decrement H$\alpha$/H$\beta$ $>5.16$ (the 90th percentile of the distribution), because dust can preferentially obscure the light from O/B stars and mimic post-starburst spectral features in star-forming spectra. This mostly excludes spaxels in the central regions of galaxies, and preferentially masks spaxels classified as star forming. Where we cannot measure the Balmer decrement due to low S/N in either the \halpha\, or \hb\, line, we assume there is no dust and perform no masking or dust correction to emission lines. Regions with no measurable Balmer decrement are typically spaxels classified as quiescent using our continuum spectral indices.
The Voronoi binned spaxels have a mean area of 1.9\,kpc$^2$ and a median of 0.1\,kpc$^2$. Our sample comprises 980000 spectra from 2404 galaxies. 
The 0.5\arcsec\, spaxel size is considerably smaller than the 2.5\arcsec\, PSF, which means that adjacent spaxels are correlated. This is somewhat mitigated by excluding small galaxies with areas similar to the PSF area, and the effect of correlated spaxels is discussed in Sections~\ref{sec:Radial_binning} and \ref{sec:Results}. At the redshift extent of our sample ($0.01<z<0.15$) the PSF size corresponds to a scale of 0.2--5.6\,kpc.

\subsection{Radial binning}
\label{sec:Radial_binning}
In order to measure the radial variation of physical properties of our galaxy sample, we split each galaxy into four elliptical radial bins. We use the Petrosian axial ratio $b/a$ and position angle from the NSA catalogue used to align the annuli with the shape of each galaxy. The major axes of the elliptical annuli are centred on 0.16, 0.41, 0.71 and 1.14 times the Petrosian effective radius, so that large and small galaxies are given equal weight in the analysis. There are 78497, 167997, 245329 and 408056 spaxels per bin, respectively. The bins are chosen to approximately equally sample the space in $R/R_e$, out to $1R_e$, with the outermost bin being wider ($1-1.5R_e$) to accommodate the larger Voronoi bin sizes typically found in the outer regions of galaxies where the SNR is lower. The spacing between the radial bins is on average smaller than the 2.5\arcsec{} PSF, so spaxels in adjacent radial bins are correlated. We verified that our conclusions are unchanged if we consider bins more than 2.5\arcsec{} apart.
Bins are allocated to radial bins if the central spaxel lies in that bin. For large Voronoi bins, it is possible that part of it lies outside of a radial bin, but in general Voronoi bins should be encompassed by the radial bin. We repeated all of our results using subsets of galaxies in different redshift ranges, and we found that our results are unchanged. This means that changes in spatial resolution with redshift, and the ability to classify galaxies by morphology due to surface brightness effects do not affect our results.

We classify all Voronoi binned spaxels in each radial bin into our five PCA classes of star-forming, quiescent, green valley, starburst and post-starburst. We also calculate the physical properties of each spaxel and fraction of spaxels in each PCA class using the total number of successfully classified spaxels in each radial bin.

\section{Which galaxies are building mass, and where?}
Observational evidence indicates that the spatial distribution of stellar mass build-up is related to the physical properties of galaxies, such as their stellar mass and morphology. High mass disc galaxies in general exhibit inside-out growth, but there is a mass and morphology dependence which is less well understood. Furthermore, galaxies which are out of dynamical equilibrium, e.g. following a merger, may be building up their mass in different locations, and at different rates, compared to galaxies which are not disturbed. We now explore where galaxies are building up their stellar mass as a function of their global properties.
To do this we first study the fraction of spaxels in each PCA class (quiescent, green valley, starburst, post-starburst, star-forming), to understand which galaxies are building mass. We then study the radial variation of the fraction of spaxels in each PCA class, to understand where the galaxies are building mass, in terms of stellar mass, concentration, and asymmetry.

\subsection{Which galaxies are building mass?}
\label{sec:Results}
Here we split our galaxy sample by stellar mass, concentration and asymmetry to find out where star-forming and quiescent spaxels are most prevalent. We calculate spaxel fraction by summing over all classified spaxels in each radial and stellar mass bin, and then further split the sample by asymmetry and concentration of the galaxy.

\subsection*{Stellar mass}
\begin{figure*}
\includegraphics[width=\textwidth]{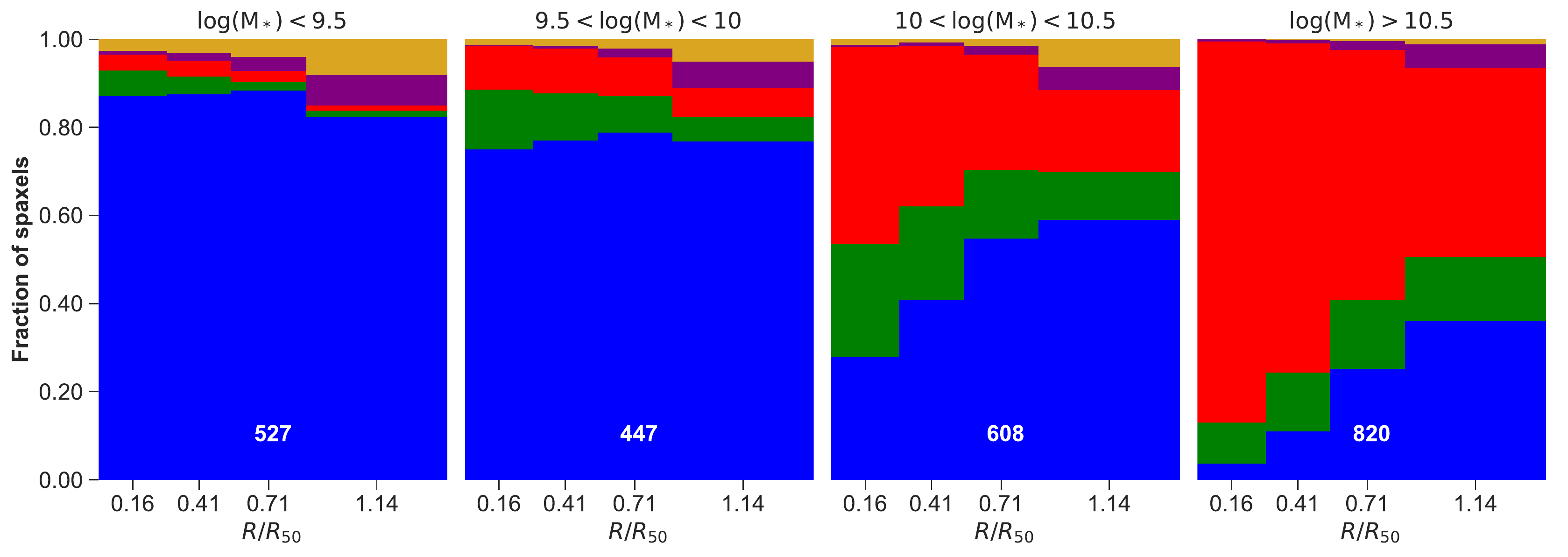}
\caption{The radial variation of each PCA class as a function of stellar mass. Each histogram shows the pixel fraction in a radial bin. The classes are: quiescent (red), star forming (blue), green valley (green), starburst (yellow) and post-starburst (purple). The number of galaxies in each bin is shown at the bottom of each panel.}
\label{fig:spaxels_Mstar}
\end{figure*}

The spaxel fraction in each radial and mass bin is shown in Figure~\ref{fig:spaxels_Mstar}.
As expected, there are clear trends as a function of mass. We find that lower mass galaxies have more star-forming, starburst and post-starburst spaxels, and the fraction drops with increasing mass. Conversely, massive galaxies ($\mstar \gtrsim 10^{10}$\msun) have more quiescent and green-valley spaxels and fewer star-forming spaxels overall. Our results are unchanged if we vary the stellar mass bin sizes and boundaries by 0.2\,dex.

\subsection*{Concentration}
In the local Universe, the high stellar mass regime is dominated by early-type galaxies, and the lower mass regime by disc galaxies. We therefore investigate trends with morphology at fixed stellar mass to disentangle the effects of these two parameters on the spatial distribution of star-forming and quiescent spaxels.

\begin{figure}
   	\includegraphics[width=0.48\textwidth]{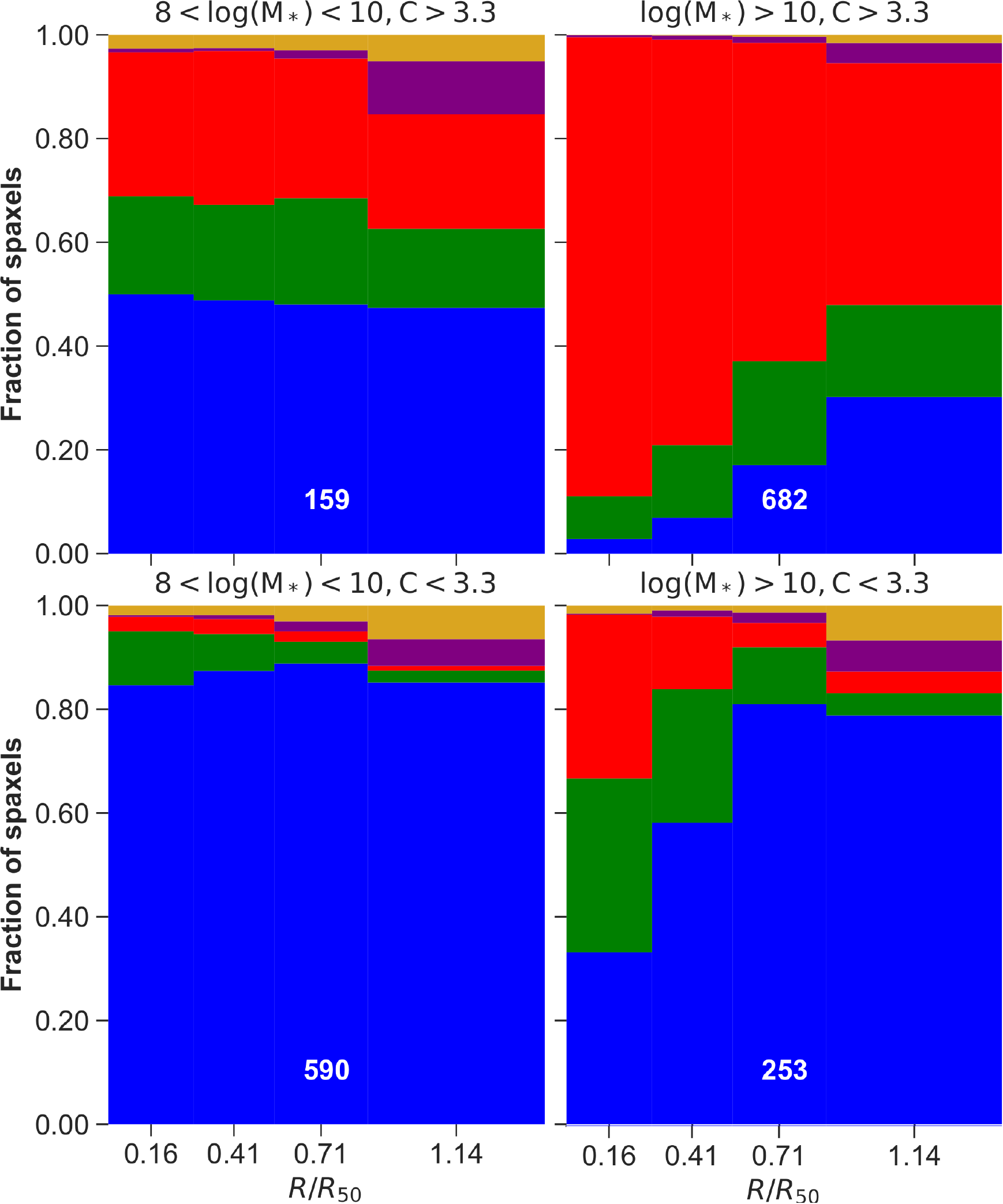}
    \caption{The radial variation of each PCA class as a function of stellar mass and concentration. Low/high values of $C$ refers to low/high concentration. Each histogram shows the pixel fraction in a radial bin. The classes are: quiescent (red), star forming (blue), green valley (green), starburst (yellow) and post-starburst (purple). The number of galaxies in each bin is shown at the bottom of each panel.}
    \label{fig:Conc_Mstar}
\end{figure}

In Figure~\ref{fig:Conc_Mstar} we show the radial variation of the fraction of spaxels in each PCA class as a function of stellar mass and concentration. Here we use concentration as a proxy for morphology: galaxies with high concentration ($C>3.3$) tend to have early-type morphologies and large bulges, and late-type galaxies tend to have low concentration indices ($C<3.3$) and therefore small bulges. In Figure~\ref{fig:Conc_Mstar_examples} we show examples of galaxies which fall in each concentration and mass bin. We now use two mass bins in order to have sufficient numbers of rare starburst and post-starburst spaxels in each mass and concentration bin. We see that at fixed stellar mass, galaxies with high concentration ($C>3.3$) have fewer star-forming and starburst spaxels, and more quiescent and green-valley spaxels. This shows us that morphology plays a significant role in predicting which galaxies have a high fraction of quiescent or star-forming spaxels, even at fixed stellar mass.

\subsection*{Asymmetry}
Asymmetric galaxies have disturbed morphologies and are either interacting or are in the pre-, ongoing or post-merger stages \citep{Lotz08, Lotz11}. Asymmetric galaxies have been found to have different physical properties compared to the general galaxy population, e.g. galaxies with high asymmetry ($A>0.2$) have bluer broadband colours \citep{Conselice99} indicating recent star formation. Galaxies with high lopsidedness (measured using the $m = 1$ azimuthal Fourier mode between the 50\% and 90\% light radii) in the outer regions tend to have low stellar mass and low surface mass density \citep{Reichard08, Reichard09}.
Most studies of asymmetric galaxies have focused on their global stellar and gas properties, or have used fibre-based measurements which can lead to aperture bias. The spatial variation of stellar populations as a function of mass and asymmetry has not yet been studied for large samples of galaxies including low mass galaxies with \mstar$<1\times10^{10}$\msun.

\begin{figure}
	\includegraphics[width=0.48\textwidth]{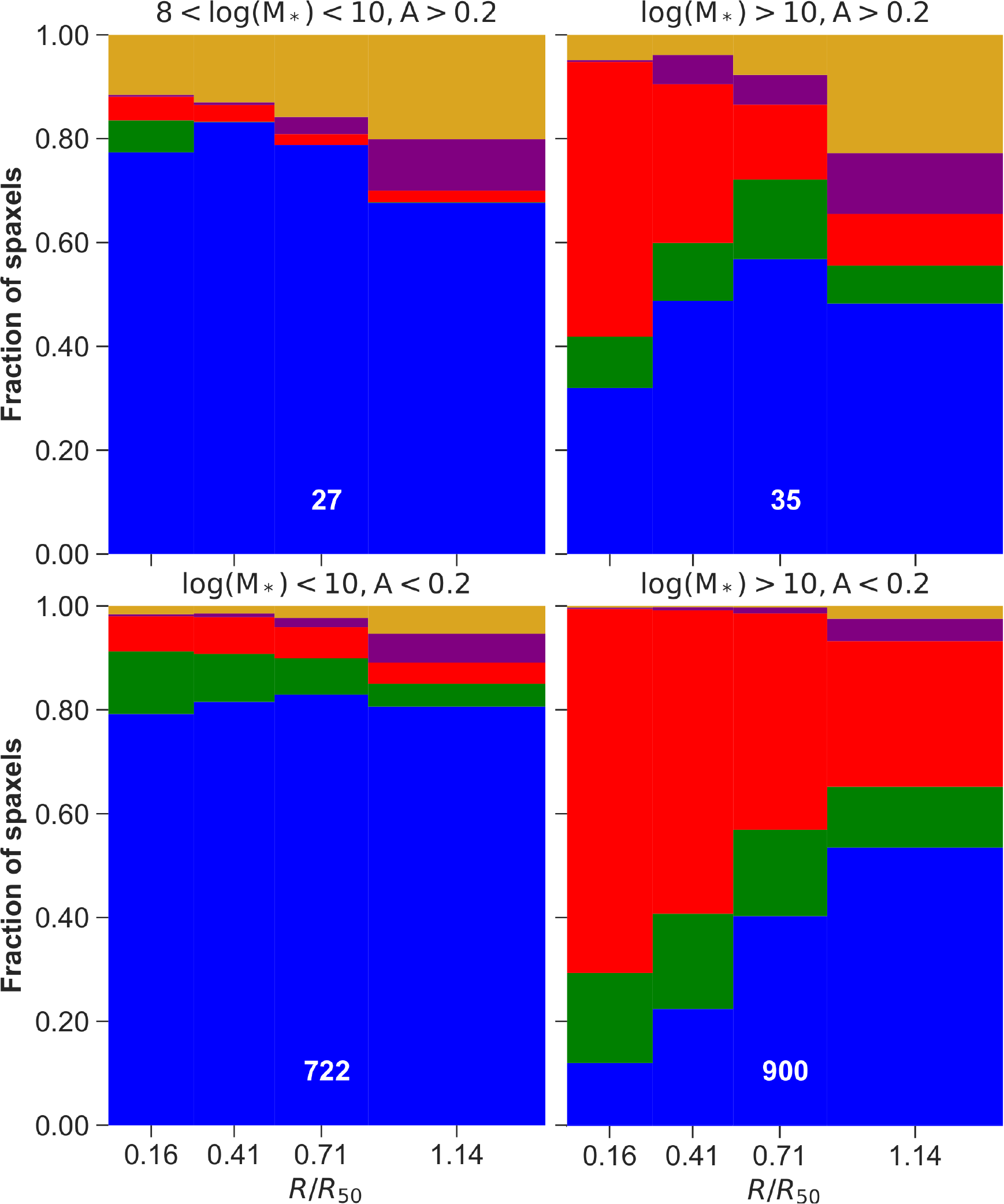}
    \caption{The radial variation of each PCA class as a function of global galaxy properties - their stellar mass and asymmetry. Low/high values of $A$ refers to low/high asymmetry. Each histogram shows the pixel fraction in a radial bin. The classes are: quiescent (red), star forming (blue), green valley (green), starburst (yellow) and post-starburst (purple). The number of galaxies in each bin is shown at the bottom of each panel.}
    \label{fig:Asym_Mstar}
\end{figure}

In Figure~\ref{fig:Asym_Mstar} we show the radial variation of the fraction of spaxels in each PCA class as a function of stellar mass and asymmetry. Asymmetric galaxies (top row of panels, A$>0.2$) have a higher fraction of starburst and post-starburst spaxels. At low mass the fraction of quiescent and green-valley spaxels are correspondingly reduced overall, and at high mass the fraction of star-forming spaxels is slightly reduced. In Figure~\ref{fig:Asym_Mstar_examples} we show examples of galaxies which fall in each asymmetry and mass bin. These results show that morphological disturbance plays a significant role in predicting where bursty star formation occurs, regardless of stellar mass. This is consistent with the overall bluer colours reported previously for asymmetric galaxies \citep{Conselice99}.

\subsection*{Other parameters}
We also examined the spaxel fraction radial profiles binned by other morphological parameters such as clumpiness ($S$, the residual of a smoothed and unsmoothed galaxy image), Gini coefficient ($G$, the relative distribution of the galaxy pixel flux values),  $M_{20}$ index (the second-order moment of the brightest 20\% of a galaxy's flux), and shape asymmetry (the asymmetry of the binary detection mask), see \citet{Lotz04} and \citep{Pawlik16}.

The clumpiness quantifies the amount of small scale structure in a galaxy, with high clumpiness values indicative of star-forming regions which are typically found in late-type galaxies. The Gini coefficient is similar to the concentration, and a high (low) value is characteristic of late-type/disturbed (early-type) galaxies, whereas the $M_{20}$ coefficient is similar to the asymmetry coefficient, with high (low) values characteristic of late-type/disturbed (early-type) galaxies. A galaxy's position on the combined Gini-$M_{20}$ plane can be used to separate early-type, late-type and morphologically disturbed galaxies\citep{Lotz04}. The shape asymmetry parameter is more sensitive to faint tidal features than the standard asymmetry measurement \citep{Pawlik16}.

We find similar trends when splitting the sample by clumpiness, Gini and M20 to those found when splitting by concentration, this is because at fixed stellar mass we are broadly splitting our sample into disc and spheroids no matter what morphological parameter we use. Similarly, using shape asymmetry and the Gini-$M_{20}$ plane to identify mergers and non-mergers produces similar trends to those seen when splitting by asymmetry.

\subsection{Where are galaxies building mass?}
We now turn to the radial distributions presented in Figures \ref{fig:spaxels_Mstar}, \ref{fig:Conc_Mstar} and \ref{fig:Asym_Mstar}.
Figure \ref{fig:spaxels_Mstar} shows that on average, high mass galaxies are building mass via star formation more at $1 R_e$ than in their centres. Low mass galaxies are building mass at all radii. The bimodality in the radial distribution of star-forming and quiescent spaxels occurs irrespective of concentration (Figure~\ref{fig:Conc_Mstar}), and asymmetry (Figure~\ref{fig:Asym_Mstar}). This shows that stellar mass correlates with the radial distribution of star-forming and quiescent spaxels to a greater extent than morphology does i.e. high mass disky galaxies are building mass predominantly at large radii, whereas low mass discs are building mass throughout; and low mass bulge dominated galaxies are building mass at all radii, whereas high mass bulge dominated galaxies are building mass predominantly in the outskirts.

On the other hand, galaxy asymmetry is clearly related to the radial distribution of bursty mass buildup.
In both low and high-mass asymmetric galaxies, starburst and post-starburst spaxels are preferentially found in the outskirts at $\sim1 R_e$. At low (high) mass there are correspondingly fewer star-forming (quiescent) spaxels. This suggests that in the outer regions of these galaxies the mode of star formation has changed from constant to bursty. 
Bursts of star-formation may have different triggers - gas accretion or mergers can deliver gas to a galaxy, disk instabilities may equally cause bursty star formation. Gas accretion may cause bursts in the outer regions of galaxies, mergers may cause funnelling of gas to central starbursts. Disc instabilities may equally be the cause of bursts in the inner parts of galaxies. In the next Section we examine the spatially resolved physical properties of the gas to work out why some regions are undergoing starbursts, while others are quiescent or transitioning.

\section{Discussion}

We have investigated the spatially resolved SFHs of galaxies and their connection to stellar mass and morphology. This allows us to see where and in what types of galaxy mass is being built by star formation, and whether the star formation is continuous or bursty.

While it is expected that there would be more star-forming spaxels in low mass galaxies, as the low mass galaxy population is dominated by star-forming galaxies \citep[e.g.][]{Baldry04}, it has become clear from recent IFU studies that high mass galaxies are not exclusively quiescent. We find that massive galaxies ($\mstar \gtrsim 10^{10}$\msun) have more quiescent spaxels in their centres (inside-out quenching) and are preferentially building mass at larger radius (inside-out growth). Low mass galaxies have a constant fraction of star-forming spaxels with radius. This mass dependence in the radial distribution of star formation agrees with many previous studies, including for example \citet{Perez13} who used CALIFA data to recover the spatially resolved SFHs of 105 galaxies.

We find a noticeable change in the radial distribution at $\mstar \sim 10^{10}$\msun, agreeing with the value found using broadband colour gradients in \citet{Pan15} and spectral indices in \citet{Wang18}.
Our method does not quantify the rate at which mass is being built by star formation, therefore we are unable to differentiate between inside-out or outside-in growth for low mass galaxies. This point was also made by \citet{Wang18} who used traditional spectral indices to trace the recent SFH in MaNGA galaxies. They found no/weak radial gradients in tracers of recent star formation in low mass galaxies, and were unable support either the ``inside-out'', or the ``outside-in'' picture.

We find that more highly concentrated galaxies have a higher fraction of quiescent spaxels, at fixed stellar mass. However, galaxies with high concentration (i.e. early-type galaxies) are not exclusively quiescent. This agrees with the low levels of star formation found in many early-type galaxies using UV-optical colours \citep{Yi05, Kaviraj07, Kaviraj08, Schawinski07}, which has been attributed to minor mergers \citep{Kaviraj09}. Our results are consistent with \citet{Spindler18} who examined the SSFR radial profiles as a function of stellar mass, S\'ersic index and central velocity dispersion. They found that intermediate and high mass galaxies with high velocity dispersion or S\'ersic index (i.e. high bulge strength) had centrally suppressed star-formation. Our results are also in agreement with \citet{GonzalezDelgado15}, who used CALIFA data to show that the centres of galaxies of earlier morphological type have older light-weighted ages. However, our results disagree with \citet{GonzalezDelgado15} in an important way: they concluded that the stellar population age and its radial variation are more closely linked to morphology than stellar mass. We clearly find that whether a spaxel is star forming or not depends on stellar mass and radius, before morphology. Presumably the difference originates in the methods used, highlighting the extreme care needed in interpreting observations of the spatial distribution of mass growth.

As for the distribution of bursty star formation, we find (post-)starburst spaxels in both low and high mass galaxies, and their fraction increases with radius. Previous results from single fibre observations suggested that low mass galaxies had a higher incidence of starburst and post-starburst spaxels \citep{Kauffmann03a}, which we find to be true at all radii. These results are in agreement with \citet{Huang13} who found that episodic star formation occurs in the outer regions of massive galaxies ($\mstar>10^{10}$\msun), ascribing this to recent gas accretion. Our study extends to lower stellar masses, and we find an equal increase in the fraction of (post-)starburst spaxels with radius at all stellar masses. This disagrees with previous results which suggested that only low-mass galaxies have bursty SFHs \citep{Bauer13, Ibarra-Medel16}.

In the following subsections we investigate possible causes for the different modes of star formation, as well as the dramatic change in the distribution of star-formation in galaxies as a function of mass and morphology.

\subsection{What causes the different modes of star formation?}

We find that the fraction of starburst and star-forming spaxels and their location in a galaxy is dependent on the asymmetry. The higher fraction of (post-)starburst spaxels in asymmetric galaxies seems qualitatively consistent with mergers/interactions triggering starbursts, leading to higher levels of star formation compared to galaxies with low asymmetry. The higher prevalence of starburst spaxels in lower mass asymmetric galaxies compared to higher mass galaxies may be due to their higher gas fractions \citep{Geha06, Bradford15}, so a merger can more easily trigger a starburst.

The canonical picture from simulations is that mergers funnel gas towards the centres of galaxies, triggering a central starburst because the gas surface density is higher \citep[e.g.][]{Bekki_Couch11, Ellison13, Moreno15}. This is in contrast to our observations where we find the enhancement of star-formation to be more prevalent in the outskirts of galaxies. We note, however, that this may be due to the rapid decline in asymmetry following a merger \citep{Pawlik18,Lotz08}. Late-stage mergers with bursts in the centre may be not be included in the high asymmetry ($A>0.2$) bins in Figure~\ref{fig:Asym_Mstar}.
Different ways of identifying mergers would be needed to explore the connection between the spatial distribution of star-formation and merger properties (such as stage and mass ratio).

To further investigate potential causes for the different modes of star formation, we must turn to the gas properties. Cold gas (atomic and molecular) is the key diagnostic of galaxy growth because gas is the fuel for star formation. Whilst we do not have resolved cold gas measurements for large numbers of MaNGA galaxies, we can infer the cold gas content using an empirical calibration between observed attenuation and gas surface density, following \citet{Barrera-Ballesteros18}. The gas-phase metallicity traces the enrichment of the interstellar medium (ISM) by previous generations of stars, and the effect of inflows and outflows of gas. Departures from the mass-metallicity or gas fraction-metallicity relation are therefore indicative of the addition of low metallicity gas to galaxies.

\begin{figure}
\includegraphics[width=0.48\textwidth]{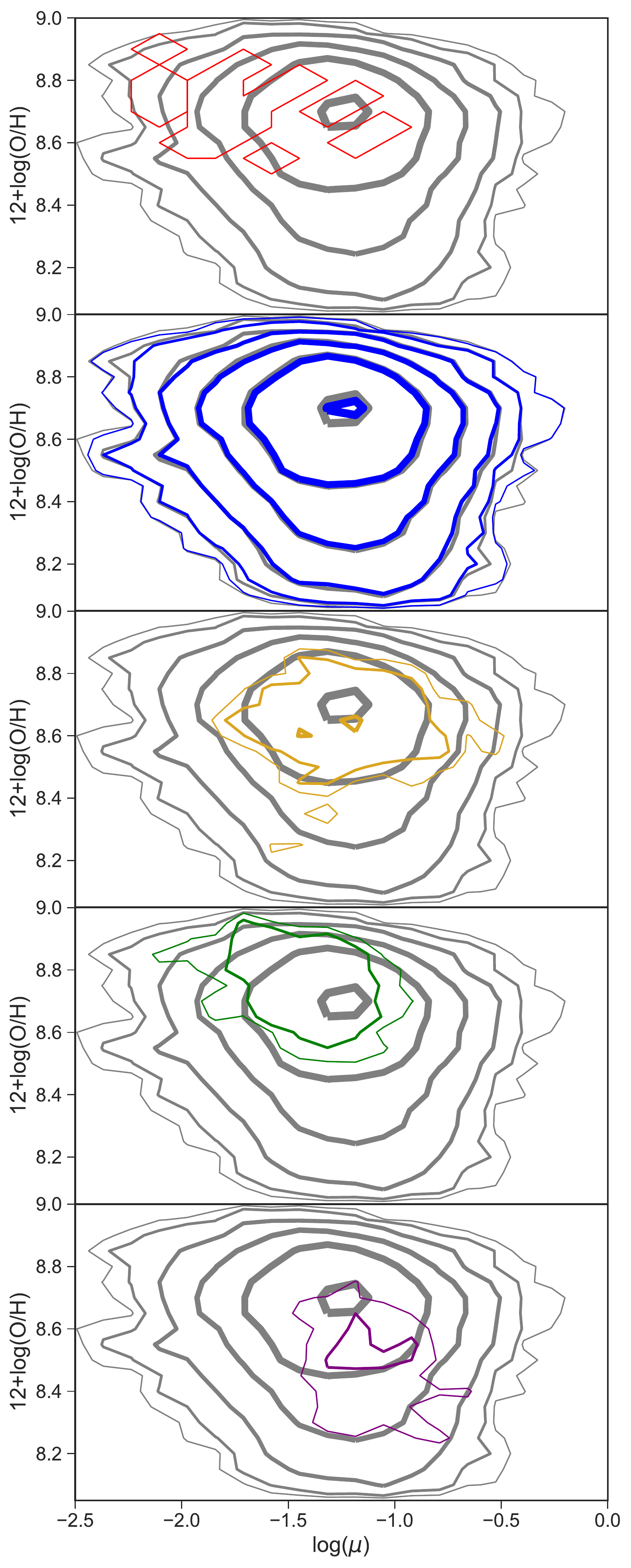}
\caption{The gas fraction ( $\mu = \Sigma_\mathrm{gas} / (\Sigma_\mathrm{gas} + \Sigma_\mathrm{*})$) and gas-phase metallicity ($12+\mathrm{log(O/H)}$) for all star-forming spaxels (black contours), split by PCA class. The classes are: quiescent (red), star forming (blue), green valley (green), starburst (yellow) and post-starburst (purple). 
Contour levels are 0.5, 1.5, 3, 10, 30, 50 and 90\% of the maximum number of of spaxels and the width corresponds to the contour level.}
\label{fig:Sigma_gas_Z}
\end{figure}

In Figure~\ref{fig:Sigma_gas_Z} we plot the gas fraction ($\mu = (\Sigma_\mathrm{gas} / (\Sigma_\mathrm{gas} + \Sigma_\mathrm{*})$) and gas-phase metallicity (using the O2N2 indicator) for all spaxels in each PCA class. Note that in order to get reliable metallicities, we only include spaxels which have been classified as star-forming in the [NII] BPT diagram \citep{BPT81, VO87}, and which have \halpha\, EW $>6$\AA. The requirement of good emission line  detections means that we exclude $\sim 50\%$ spaxels without emission from this part of the analysis.
This should not bias the starburst and star-forming samples since these should have strong emission lines. Quiescent, green valley and post-starburst samples  may be biased towards a particular subset of the population (e.g. having low-level star-formation).

The star-forming spaxels have typical gas fractions of $\sim 0.1$, which is consistent with the value found for normal nearby galaxies \citep{Saintonge11a}. The gas-phase metallicities of the star-forming spaxels are typically $12+\mathrm{log(O/H)}=8.7$, with a tail to lower metallicities. There are very few quiescent and green valley spaxels because we have excluded spaxels with non-star-forming emission line ratios. Those quiescent spaxels which are retained in the sample have low gas fractions and high gas-phase metallicities. The few green valley spaxels have low gas fraction but perhaps slightly higher metallicities.

Now we examine the spaxels with evidence of bursty SFH. Those post-starburst spaxels retained in the sample have a slightly higher gas fraction on average (0.1\,dex) compared to star-forming spaxels, a peak metallicity around 8.6 with a tail to lower metallicities. The metallicity of the starburst regions has a bimodal distribution with peaks at $12+\mathrm{log(O/H)}=8.7$ and 8.4. Starburst spaxels have on average a slightly higher gas fraction compared to star-forming spaxels (0.2\,dex). This shows that bursty star formation  is occurring preferentially in regions with higher gas content than star-forming spaxels. The bimodal distribution of metallicities of starburst spaxels may indicate that some bursts are triggered by inflows or accretion of lower metallicity gas.

\subsubsection{Gas fraction radial profiles}
\label{sec:gas_Z_radial}

\begin{figure}
	\includegraphics[width=0.48\textwidth]{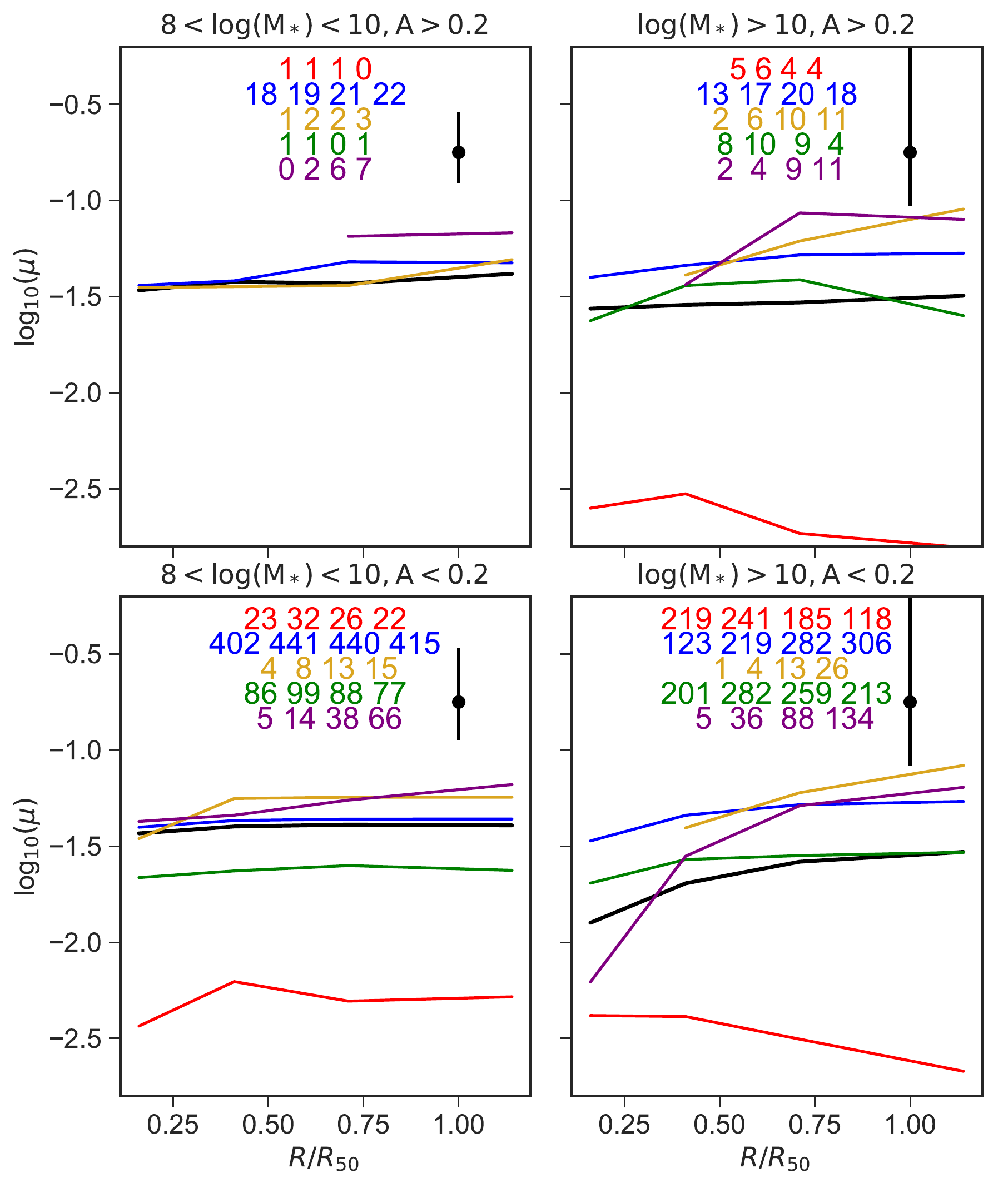}
   \caption{The median radial variation of the gas fraction ($\mu = \Sigma_\mathrm{gas} / (\Sigma_\mathrm{gas} + \Sigma_\mathrm{*})$) for spaxels in each PCA class as a function of stellar mass and asymmetry. The classes are: all (black), quiescent (red), star forming (blue), green valley (green), starburst (yellow) and post-starburst (purple). The number of galaxies in each radial bin and class in shown at the top of each panel. Lines are not shown where there are less than three galaxies contributing to a radial bin. The black error bar shows the 16th--84th percentile range of the total sample, averaged over all radial bins.
   The typical uncertainty on the gas fraction is a factor of two.}
   \label{fig:Asym_gasfrac}
\end{figure}

\citet{Moran12} found that galaxies with the largest metallicity drops in their outer regions have high HI content, indicating that inflows of gas are related to low metallicity. Furthermore, galaxies with a large metallicity drop were found to be building stellar mass at a higher rate than other galaxies in their sample. If bursts are triggered by accretion of gas in asymmetric galaxies by mergers, then we may expect to see a lower metallicity than expected in certain regions of a galaxy.
We first examine the median radial variation of spaxel gas fraction for each PCA class in Figure~\ref{fig:Asym_gasfrac}. To do this we calculate the mean gas fraction of spaxels in each PCA class in four elliptical radial bins for each galaxy. We then take the median of the gas fraction in each PCA class for the galaxy sample to produce radial profiles of gas fraction in mass and asymmetry bins.

We find that gas fraction increases from the inner to outer regions for all galaxies on average, regardless of PCA class. Star-forming spaxels have significantly higher gas fractions than green valley and quiescent spaxels in all mass and asymmetry bins. The starburst spaxels have similar gas fractions to star-forming spaxels, except at large radii: the outer starburst and post-starburst regions have slightly higher gas fractions than the star-forming spaxels. However, we find that this is not because there is more gas, but because the stellar mass density decreases to larger radius. This suggests that gas fraction does not determine whether star formation is bursty, in low or high mass galaxies, asymmetric or symmetric, even at large radius.

\subsubsection{Metallicity radial profiles}
We can use the spatially resolved information from MaNGA to explore the relation between gas-phase metallicity and SFH in asymmetric and non-asymmetric galaxies. Using metallicity gradients, \citet{Moran12} found that the gas-phase metallicity in the outer regions of star forming galaxies decreases as total atomic gas fraction increases, which may be evidence of accretion of gas onto the outer regions of galaxies. The relation between gas-phase metallicity and gas fraction is an important tool in understanding galaxy evolution.

\begin{figure}
	\includegraphics[width=0.48\textwidth]{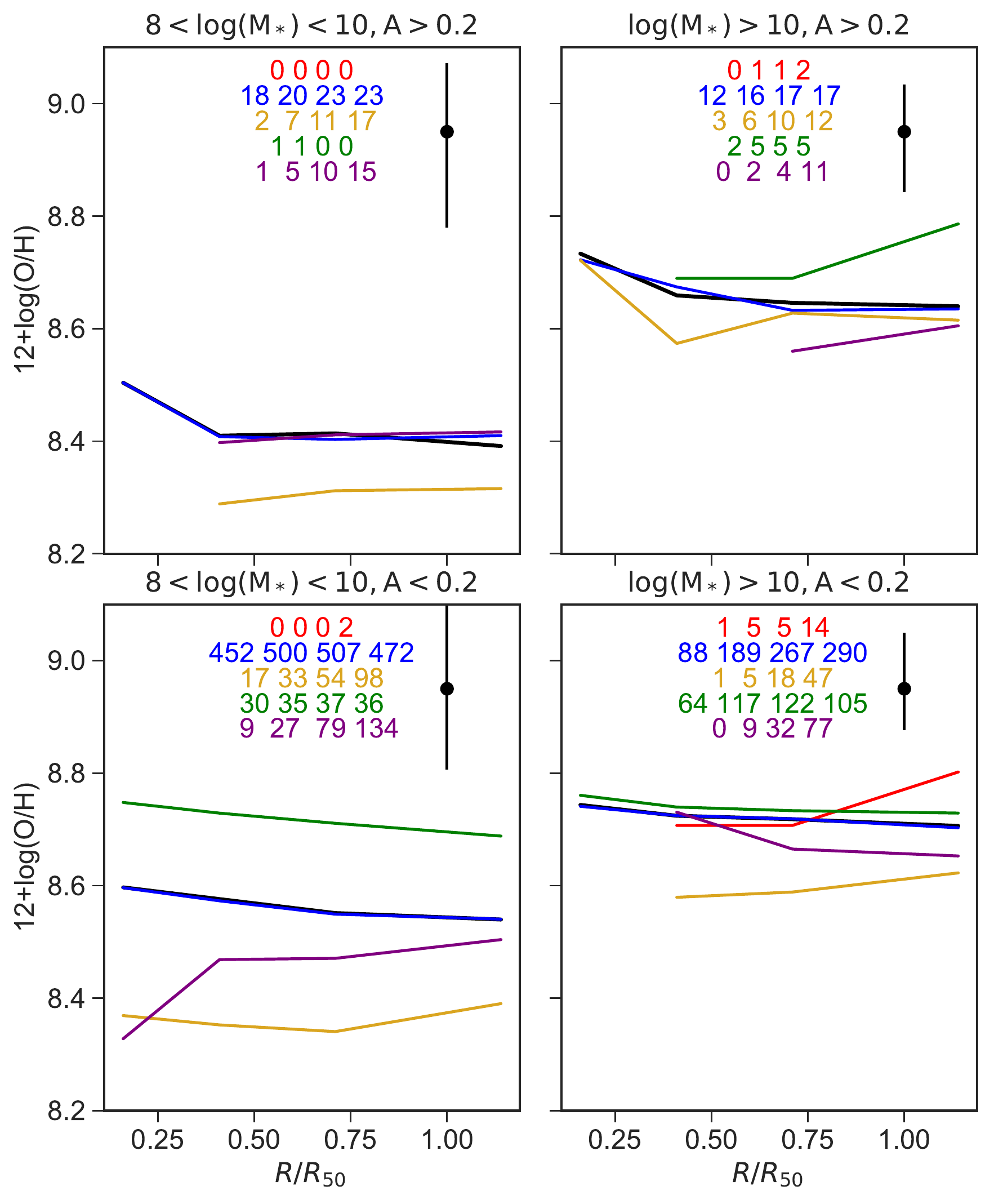}
   \caption{The median radial variation of the gas phase metallicity of spaxels in each PCA class as a function of stellar mass and asymmetry. The classes are: all (black), quiescent (red), star forming (blue), green valley (green), starburst (yellow) and post-starburst (purple). There are very few quiescent spaxels. The number of galaxies in each radial bin and class in shown at the top of each panel. Lines are not shown where there are less than three galaxies contributing to a radial bin. The black error bar shows the 16th--84th percentile range of the total sample, averaged over all radial bins.
   }
   \label{fig:Asymmetallicity}
\end{figure}

In Figure~\ref{fig:Asymmetallicity} we plot the median radial variation of gas-phase metallicity of spaxels of each PCA class, in bins of mass and asymmetry. We find that the overall metallicity increases with stellar mass, as expected from the fundamental mass-metallicity relation \citep{Tremonti04}. For low mass galaxies, the metallicity of spaxels of all PCA class is slightly lower in asymmetric galaxies, at all radii. This is consistent with the accretion of gas from a merger lowering the metallicity at all radii up to $\sim 1\,R_e$, not just in the galaxy centre as has been shown in previous studies \citet{Kewley06, Ellison08, Reichard09} using single fiber spectroscopy. In high mass galaxies there is no clear difference in metallicity between asymmetric and symmetric galaxies.

\begin{figure}
     \includegraphics[width=0.48\textwidth]{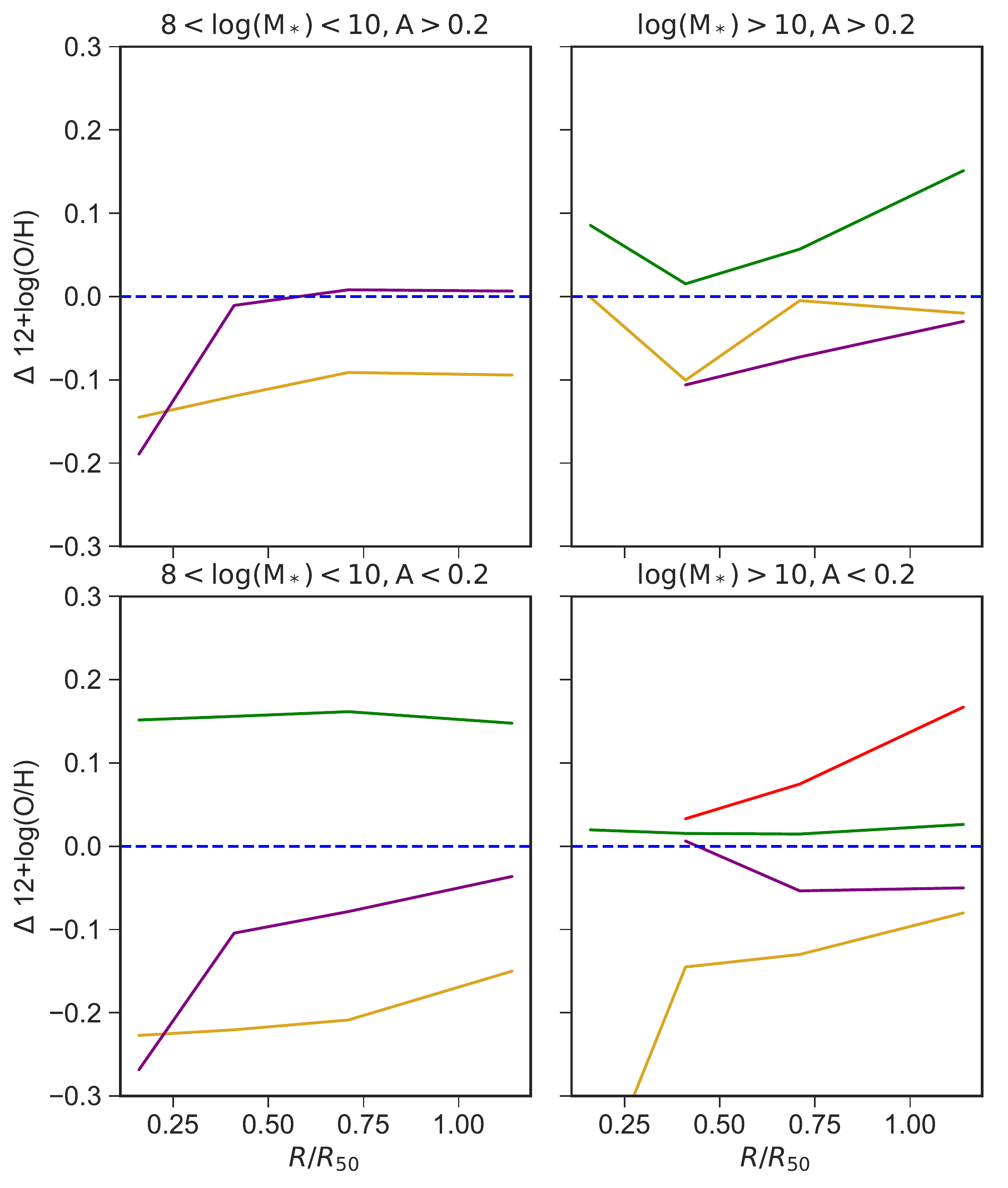}
   \caption{The radial variation of the gas phase metallicity of spaxels in each PCA class relative to star-forming spaxels. The classes are: star forming (blue), quiescent (red), green valley (green), starburst (yellow) and post-starburst (purple). The number of galaxies in each radial bin and class in shown at the top of each panel.}
   \label{fig:Asymmetallicityresid}
\end{figure}

The metallicity of regions undergoing starbursts is lower than all other PCA classes, at all radii, with the exact deficit depending on both stellar mass and asymmetry. This is explored further in Figure~\ref{fig:Asymmetallicityresid} where we plot the difference between the median radial metallicity profile of star-forming spaxels and those in other PCA classes. The deficit in metallicity of starburst spaxels relative to star-forming spaxels in the low mass bin, with $\Delta 12+\mathrm{log(O/H)} \sim 0.1-0.15$\,dex in non-asymmetric galaxies, and $\Delta 12+\mathrm{log(O/H)} \sim 0.05-0.1$\,dex in asymmetric galaxies. In high mass galaxies the deficit in metallicity of starburst spaxels is not as large as we find in low mass galaxies, particularly for asymmetric galaxies. Post-starburst spaxels typically have metallicity values closer to star-forming spaxels than starburst spaxels at all radii. This may be because the ISM has been enriched with metals following the starburst. The few green valley and quiescent spaxels with significant emission lines have higher metallicities than star-forming spaxels.

Our finding that the overall metallicity of spaxels is lower in asymmetric galaxies is qualitatively similar to \citet{Ellison08}, who found that galaxies in close pairs have a $\sim 0.05$\,dex lower central metallicity compared to field galaxies at a given stellar mass, see also \citet{Michel-Dansac08}. Post-merger galaxies show an even larger metallicity deficit ($0.1$\,dex) than close pairs \citep{Ellison13}. Larger changes in metallicity were found by \citet{Kewley06}, who found that galaxy pairs had a $0.2$\,dex lower metallicity at a given luminosity, compared to a control sample. However, \citet{Scudder12} found smaller differences in metallicity with decreasing projected separation between galaxy pairs. Our results are consistent with \citet{Reichard09}, who found using SDSS single fibre data that metal poor galaxies are also the most lopsided. 
\citet*{Peeples09} examined the properties of galaxies lying below the mass-metallicity relation and found that the majority showed signed of mergers and were extremely blue in colour, indicating high SSFR. A similar result was found in \citet{Reichard09} where lopsided galaxies had galaxy centres with $0.05-0.15$\,dex lower metallicity than non-lopsided galaxies. \citet{Lee04} found that morphologically disturbed galaxies have a lower metallicity than galaxies of similar luminosity. \citet{Gronnow15} found that the properties of galaxies below the fundamental mass-metallicity relation are consistent with a model in which a merger has recently occurred. They found that 1:5 to 1:1 mergers cause an average 0.11\,dex decrease in metallicity. Since all of these studies are based on single fiber SDSS spectra, only the central region of the galaxy was sampled.
Our study allows us to explore the metallicity in all regions of a galaxy out to $R_e \sim 1.5$, showing that the deficit in metallicity is global rather than central.

Our findings are in agreement with numerical simulations of galaxy mergers, where gas is transported inwards from the outer regions causes a dilution of the metallicity \citep{Montuori10,Rupke10, Perez11}. Whether the metallicity is ultimately diluted or enhanced in the simulations depends on the gas fraction of the progenitors and time of observation. \citet{Torrey12} found that gas-poor disc-disc interactions cause a 0.07\,dex drop in the metallicity, while gas-rich interactions can result in an overall enhancement in the central metallicity following the merger. This is due to metal enrichment of the ISM from the merger-induced central starburst, which overrides any inflow of metal poor gas. This may explain why we find that the post-starburst spaxels show an enhanced metallicity compared to the starburst spaxels.

It is likely that mergers are partly responsible for delivering low metallicity gas and triggering starbursts in galaxies. However, since we find bursts of star formation outside of the galaxy centre of non-disturbed galaxies, it is possible that smooth accretion of low metallicity gas from the circumgalactic medium is also responsible for increasing the gas density. Such conclusions were reached by \citet{Wang11}, who found that galaxies with larger HI gas fractions had more actively star-forming outer discs compared to the inner part of the galaxy. This may be responsible for the low metallicity in starburst regions in the outer parts of high mass galaxies, and at all radii in low mass galaxies.  Using 800 MaNGA galaxies, \citet{Ellison18} found that starburst galaxies above the main sequence have lower gas phase metallicities out to $2 R_e$, suggesting metal-poor gas inflows accompany starbursts in general. Further examination of the abundances in bursty regions and comparison to chemical evolution models may allow us to differentiate between the source of gas in the outer parts of galaxies, in particular in non-disturbed galaxies (Hwang et al. in prep).

\subsection{Why do we see differences in the spatially resolved SFH of galaxies?}
We find that massive ($\mstar \gtrsim 10^{10}$\msun) and more concentrated ($C>3.3$) galaxies have less star-forming spaxels overall, in particular in their centres. High mass galaxies built up their central mass earlier, and their star formation stopped earlier compared to the outer regions. There are at least three possible mechanisms suggested by simulations that may explain these differences. We will address each one in turn.

A transition stellar mass of $10^{10.5}\msun$ is predicted to be a division in the dominant feedback mechanism in galaxies \citep{Shankar06}, with AGN feedback dominating above this mass, and supernova feedback being more important at lower stellar masses (see also \citealt{Kaviraj07a}).
The different predicted efficiencies of these feedback mechanisms at stopping star formation, and their different spatial distributions for energy deposition, could give rise to differences in the fraction of quiescent spaxels and spatial distribution of star formation.

\begin{figure}
	\includegraphics[width=0.49\textwidth]{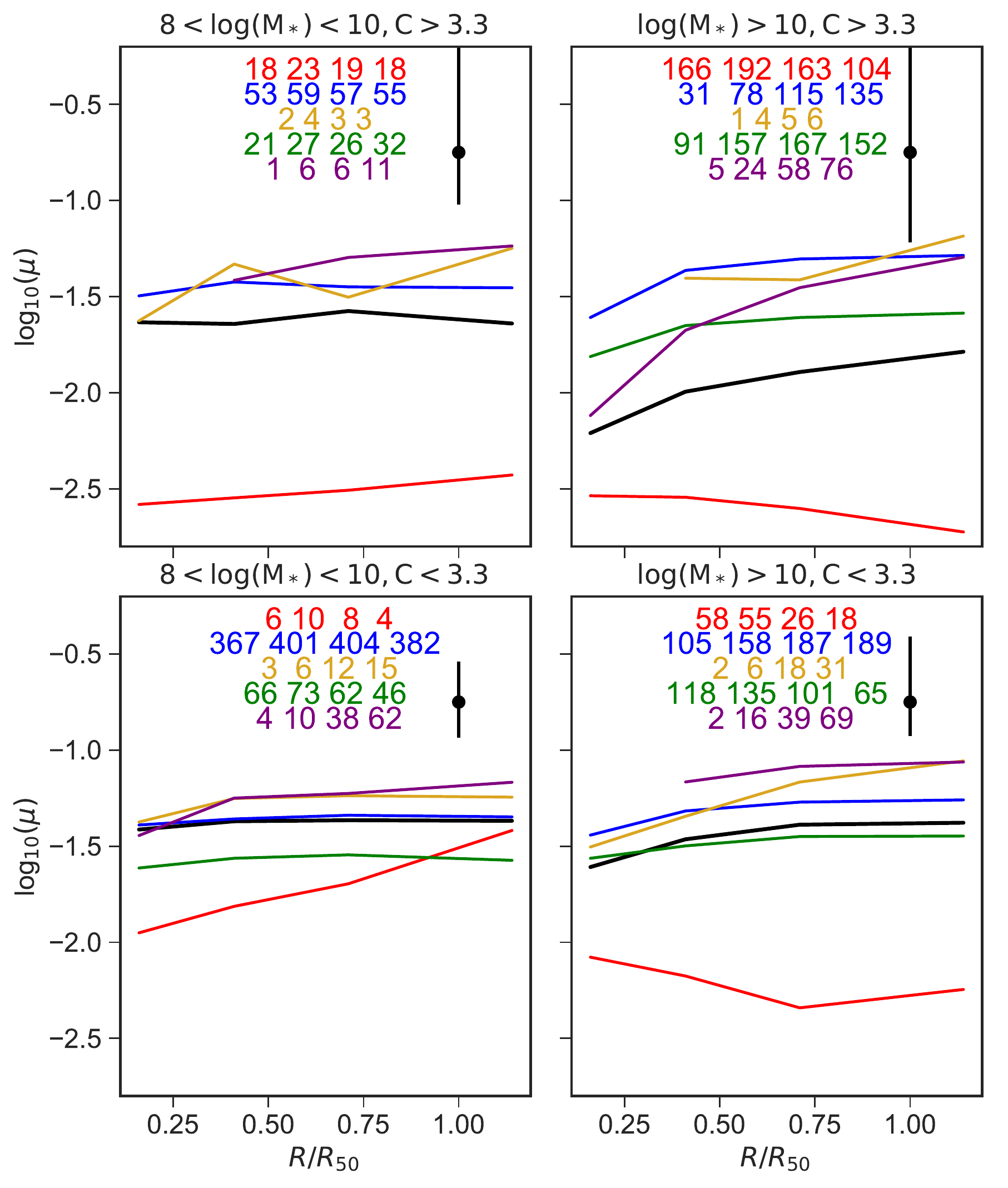}
   \caption{The median radial variation of the gas fraction ( $\mu = \Sigma_\mathrm{gas} / (\Sigma_\mathrm{gas} + \Sigma_\mathrm{*})$) for spaxels in each PCA class as a function of stellar mass and concentration. The classes are: all (black), quiescent (red), star forming (blue), green valley (green), starburst (yellow) and post-starburst (purple). The number of galaxies in each radial bin and class in shown at the top of each panel. Lines are not shown where there are less than three galaxies contributing to a radial bin. The black error bar shows the 16th--84th percentile range of the total sample, averaged over all radial bins.
   The typical uncertainty on the gas fraction is a factor of two.}
   \label{fig:Concentration_gasfrac}
\end{figure}

However, more recent simulations are able to reproduce the observed inside-out quenching of Milky Way mass galaxies without an AGN \citep{Tacchella16b, Avila-Reese18}. In these simulations galaxies undergo a central starburst at high redshift driven by a merger or disc instability \citep{Dekel_Burkert14}, which exhausts the gas supply and causes a buildup in central stellar density. The simulations of \citet{Forbes12} and \citet{Goldbaum15} showed that at $z=2$ gravitational instabilities may drive gas from the outer regions to the centre of galaxy discs, then at low redshift the gas is depleted in the centre. This is due to a combination of consumption by star formation and gravitational stabilisation of the gas as the central stellar mass density increases at late times. Galaxy centres may therefore remain quiescent if gas is no longer resupplied. In Figure~\ref{fig:Concentration_gasfrac} we  examine the median radial variation of spaxel gas fraction for each PCA class  in mass and concentration bins. We find a lower gas fraction in more concentrated galaxies, and at high masses the central regions have a much lower gas fraction compared to galaxies with low concentration. We note that observations of M$_\ast>10^{10}$\msun low redshift galaxies have found that the presence of a bulge does not strongly affect either the total molecular \citep{Saintonge11a} or total atomic hydrogen gas fraction \citep{Catinella10, Fabello11}. Clearly, spatially resolved cold gas observations of larger samples of galaxies are now required to investigate this further.

The lack of central star-formation in high mass galaxies and concentrated galaxies may be caused by morphological quenching \citep{Martig09}, whereby gravity-driven turbulence caused by large-scale, non-axisymmetric instabilities can stabilise gas against collapse. This means that a higher Jeans mass is needed in order for gas to collapse to form stars.
The higher shear induced by galactic rotation in the deep, inner part of the potential, means that gas can exist in early-type galaxies and bulges (i.e. massive and concentrated galaxies) but will not collapse to form stars \citep{Martig13, Davis14, Federrath16}.
Because the gravitational potential well is deeper in more massive galaxies, this would cause more massive and concentrated galaxies to have more quiescent spaxels. This effect is likely to be important across a range of redshifts. Morphological quenching is thought to be why some early-type galaxies in the local Universe have large gas reservoirs but are not forming stars efficiently. The process is also likely to be important at higher redshift \citep{Martig09} where gas fractions are higher and may keep newly formed quiescent galaxies from forming more stars.

Whatever internal processes are at play, the environment in which a galaxy evolves will affect its gas supply and therefore its potential to continue to form stars. Simulations of massive galaxy halos ($>10^{12}$\msun) at $z=0-3$ show infalling gas from the cosmic web shock heating to the virial temperature of the dark matter halo \citep[e.g.][]{Birnboim_Dekel03, Keres05}.
This results in a suppression of the cold gas supply to galaxy halos and is often called ``halo quenching''. The mass of $\sim 10^{12}$\msun above which this process is predicted to be active corresponds to a stellar mass of $\sim 10^{10.5}$\msun \citep{Behroozi10, Moster10}, which is around where we see the break in galaxy physical parameters \citep{Kauffmann03a, Dekel_Birnboim06, Cattaneo06}.

The radial migration of stars could play an in important role in the observed spatial distribution of stars of different ages. \citet{Avila-Reese18} used simulations of Milky-Way like galaxies to show that the radial distribution of stellar mass growth is mostly from in-situ star formation, and that radial stellar mass migration does not significantly change the distribution of stellar mass in a galaxy. However in low mass galaxies ($\mstar<10^{9.6}$\msun), inflows and outflows of gas can cause strong radial migration of stars, with young stars experiencing the strongest short-timescale migration. The oldest stars will have migrated the furthest because they have experienced more periods of gas inflow/outflow \citep{El-Badry16}. This could be an explanation for why we observe a flat distribution of star-forming and quiescent spaxels with radius in low stellar mass galaxies.

\section{Conclusions}

We have mapped the spatially resolved star-formation histories of 980000 spaxels in 2404 MaNGA galaxies. We examine the spatial distribution of star-forming, starburst, quiescent, post-starburst and green valley spaxels as a function of stellar mass and morphology. This allows us to see where and in what types of galaxy is mass being built via star formation and where quenching is occurring. We have also examined the physical properties of the ISM in different regions using the gas-phase metallicity and a proxy for the gas fraction. Our main conclusions are summarised as follows:

\begin{itemize}

\item Stellar mass affects the radial distribution of quiescent and star-forming spaxels, with a change in the distribution at $\mstar \sim 10^{10}\msun$. There are weak radial trends in the fraction of star-forming spaxels at low mass, but strong trends at high mass with a higher fraction of star-forming spaxels in the outer regions. This means low mass galaxies are building mass everywhere, while high mass galaxies are building mass in the outer regions but have largely stopped in the inner regions.

\item Galaxy concentration is related to the relative amount of star-forming and quiescent spaxels, but does not affect the radial distribution of star-forming spaxels at fixed stellar mass.

\item Asymmetry correlates with both the amount and distribution of star formation. In low (high) mass asymmetric galaxies there are fewer constant mode star-formation (quiescent) spaxels, and more starbursts and post-starburst spaxels compared to non-asymmetric galaxies. Starburst regions are more common at $1.5R_e$ than in the galaxy centre.

\item Asymmetric galaxies have lower metallicities at all radii ($<1.5R_e$), consistent with interactions triggering starbursts and driving low metallicity gas into the central regions.

\item Starburst regions have lower metallicities and similar gas fractions compared to non-starbursting regions. We find bursts of star formation outside of the galaxy centre of non-disturbed galaxies which could be fuelled by smooth accretion of low metallicity gas from the circumgalactic medium.

\end{itemize}

Our results contribute to a consistent picture in which high mass galaxies built their mass from the inside-out, and became quiescent first in the central regions. Low mass galaxies have a flatter radial distribution of stellar mass growth, with growth continuing to the present day at all radii. Given that the majority of the stellar mass in massive galaxies was formed many Gyr ago, it is hard to disentangle the exact processes leading to the growth of mass and the cause of the shutdown in star formation from the fossil record provided by local galaxies alone, since many processes will have acted upon each galaxy over cosmic time. It is likely that a combination of the ability of galaxies to accrete gas, and the regulation of star formation via feedback gives rise to the bimodality in spatially resolved SFHs observed in galaxies as a function of mass and concentration. Comparison to the spatial distribution of star-formation over time in numerical simulations would allow us to put better constraints on the processes driving mass growth and quenching.
Spatially resolved observations of cold molecular gas in these galaxies, and their progenitors at high redshift, would allow us to probe in more detail the processes regulating star-formation and galaxy growth over cosmic time.

\section*{Acknowledgements}
We thank the reviewer for their comments which helped improve the clarity of the paper.
V.~W. acknowledges support from the
European Research Council Starting Grant SEDmorph (P.I. V.~Wild). N.~L.~Z. is supported by the Catalyst award made by the Johns Hopkins University.

Funding for the Sloan Digital Sky Survey IV has been provided by the
Alfred P. Sloan Foundation, the U.S. Department of Energy Office of
Science, and the Participating Institutions. SDSS acknowledges
support and resources from the centre for High-Performance Computing at
the University of Utah. The SDSS web site is www.sdss.org.

SDSS is managed by the Astrophysical Research Consortium for the Participating Institutions of the SDSS Collaboration including the Brazilian Participation Group, the Carnegie Institution for Science, Carnegie Mellon University, the Chilean Participation Group, the French Participation Group, Harvard-Smithsonian centre for Astrophysics, Instituto de Astrof{\'i}sica de Canarias, The Johns Hopkins University, Kavli Institute for the Physics and Mathematics of the Universe (IPMU) / University of Tokyo, Lawrence Berkeley National Laboratory, Leibniz Institut f{\"u}r Astrophysik Potsdam (AIP), Max-Planck-Institut f{\"u}r Astronomie (MPIA Heidelberg), Max-Planck-Institut f{\"u}r Astrophysik (MPA Garching), Max-Planck-Institut f{\"u}r Extraterrestrische Physik (MPE), National Astronomical Observatories of China, New Mexico State University, New York University, University of Notre Dame, Observat{\'o}rio Nacional / MCTI, The Ohio State University, Pennsylvania State University, Shanghai Astronomical Observatory, United Kingdom Participation Group, Universidad Nacional Aut{\'o}noma de M{\'e}xico, University of Arizona, University of Colorado Boulder, University of Oxford, University of Portsmouth, University of Utah, University of Virginia, University of Washington, University of Wisconsin, Vanderbilt University, and Yale University.

This project makes use of the MaNGA-Pipe3D dataproducts. We thank the IA-UNAM MaNGA team for creating it, and the ConaCyt-180125 project for supporting them.

The Pan-STARRS1 Surveys (PS1) and the PS1 public science archive have been made possible through contributions by the Institute for Astronomy, the University of Hawaii, the Pan-STARRS Project Office, the Max-Planck Society and its participating institutes, the Max Planck Institute for Astronomy, Heidelberg and the Max Planck Institute for Extraterrestrial Physics, Garching, The Johns Hopkins University, Durham University, the University of Edinburgh, the Queen's University Belfast, the Harvard-Smithsonian centre for Astrophysics, the Las Cumbres Observatory Global Telescope Network Incorporated, the National Central University of Taiwan, the Space Telescope Science Institute, the National Aeronautics and Space Administration under Grant No. NNX08AR22G issued through the Planetary Science Division of the NASA Science Mission Directorate, the National Science Foundation Grant No. AST-1238877, the University of Maryland, Eotvos Lorand University (ELTE), the Los Alamos National Laboratory, and the Gordon and Betty Moore Foundation.

This research  has  made  use  of  NASA's  Astrophysics  Data  System Bibliographic  Services, NumPy \citep{Walt2011}, Matplotlib \citep{Hunter07}, Astropy, a community-developed core Python package for Astronomy \citep{Astropy, Astropy2}, \url{http://www.astropy.org}, the PhotUtils package \citep{Photutils} and Marvin, a core Python package and web framework for MaNGA data, developed by Brian Cherinka, Jos{\'e} S{\'a}nchez-Gallego, and Brett Andrews \citep{brian_cherinka_2017_292632}.
This research has made use of \textsc{pyqz} \citep{Dopita13}, a Python module to derive the ionization parameter and oxygen abundance of HII regions from their strong emission line ratios hosted at \url{http://fpavogt.github.io/pyqz}. \textsc{pyqz} relies on \textsc{statsmodel} \citep{seabold2010} and \textsc{matplotlib} \citep{Hunter07}.


\bibliographystyle{mnras}
\bibliography{refs}


\begin{appendix}

\section{Rejected Spectra}
\label{sec:Reject_spectra}

\begin{figure}
\includegraphics[width=0.48\textwidth]{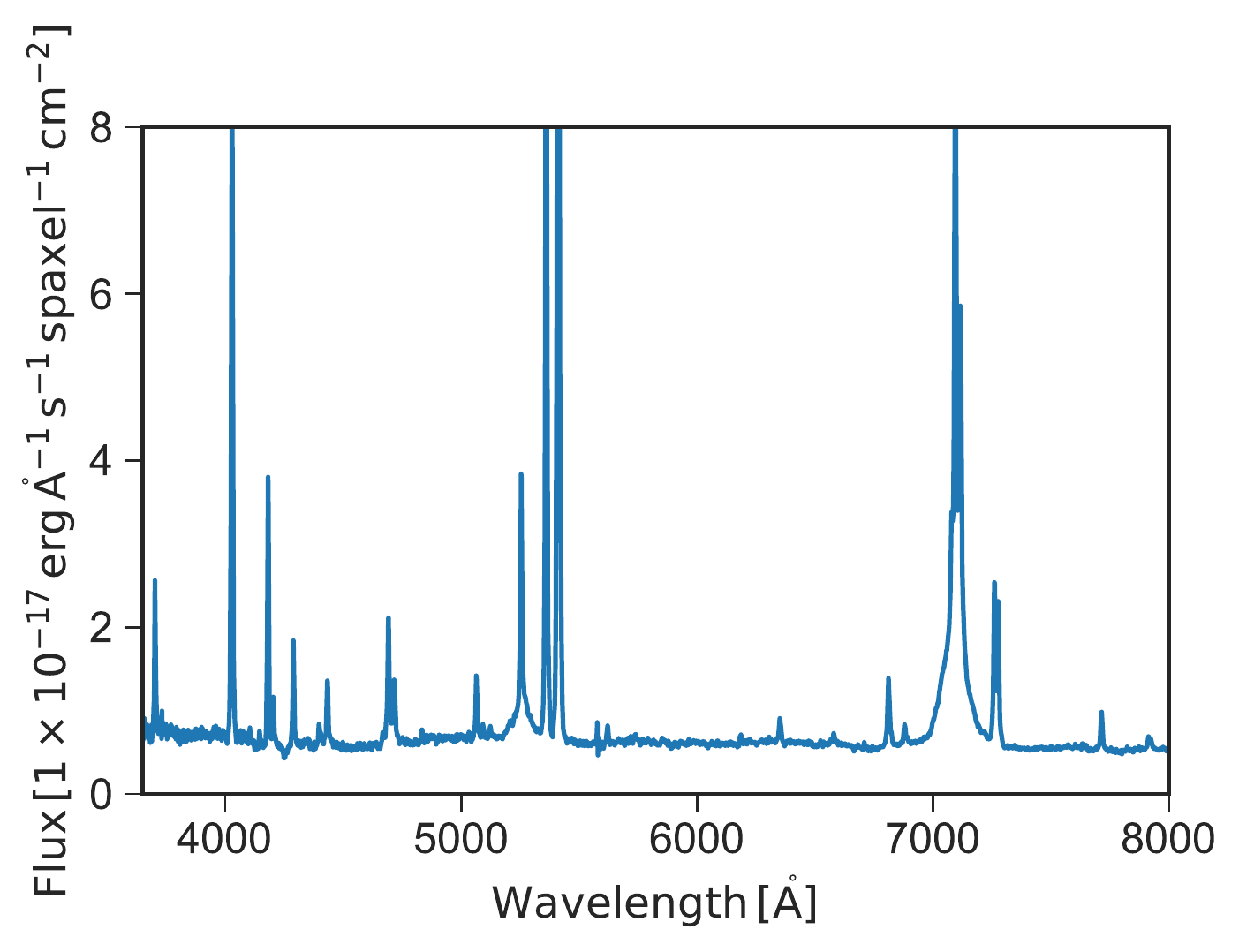}
\includegraphics[width=0.48\textwidth]{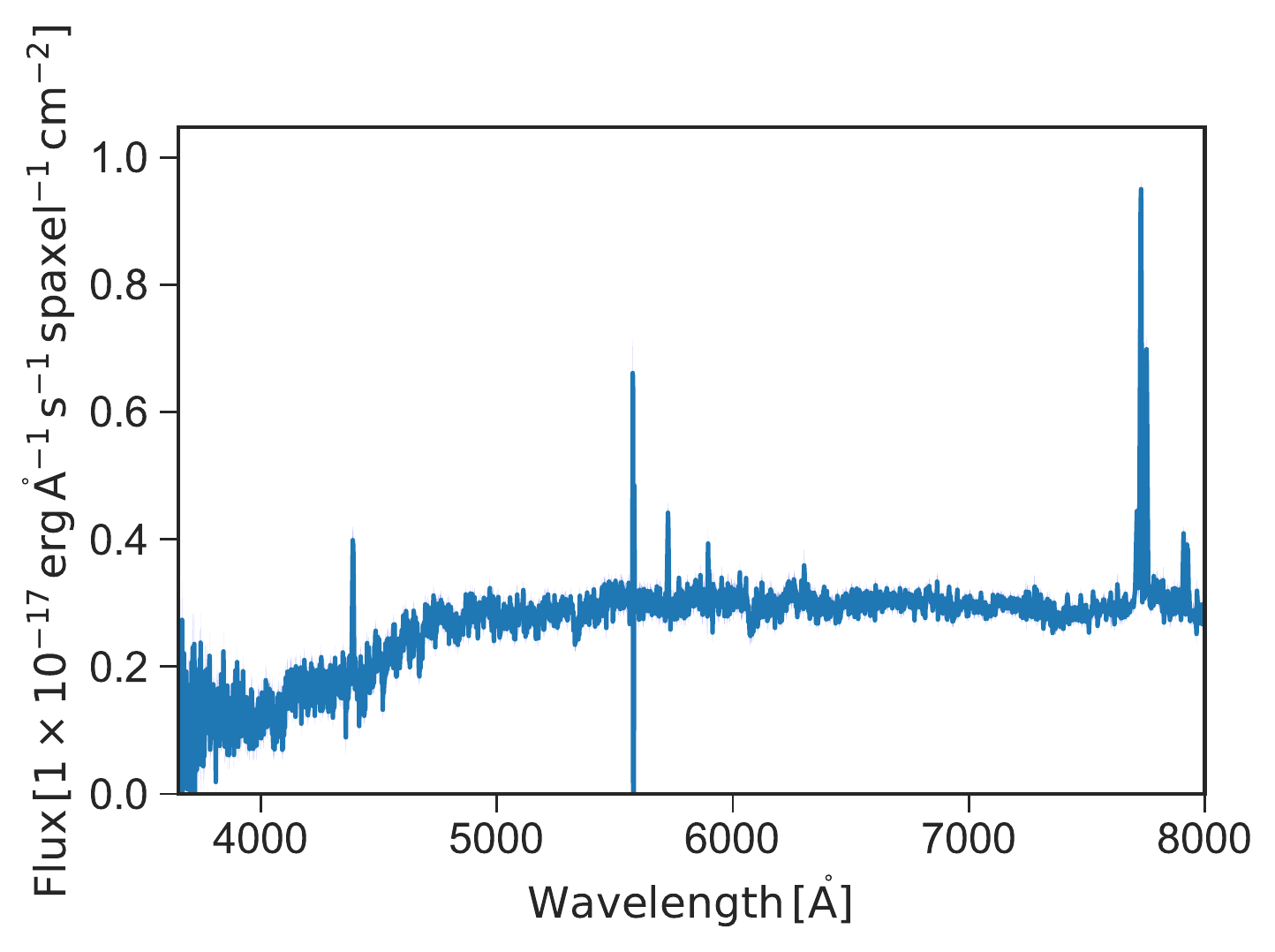}
   \caption{Example spectrum which fall below the star-forming region in PCA space. Spectra are commonly broadline AGN or are very noisy spectra.}
   \label{fig:Rejected_spectra}
\end{figure}

Points below the star-forming region are spectra with broadline AGN or have unusual continuum shapes. 
These are poorly described by the PCA reconstructions and are excluded from our analysis. In Figure~\ref{fig:Rejected_spectra} we show examples of rejected spaxels. A small number of highly star-forming spaxels are removed due by conservative cuts at low values of PC2 to exclude spaxels dominated by light from a broadline AGN.

\section{Examples of morphology classes}
\label{sec:Morph_examples}

\begin{figure*}
	\includegraphics[width=0.99\textwidth]{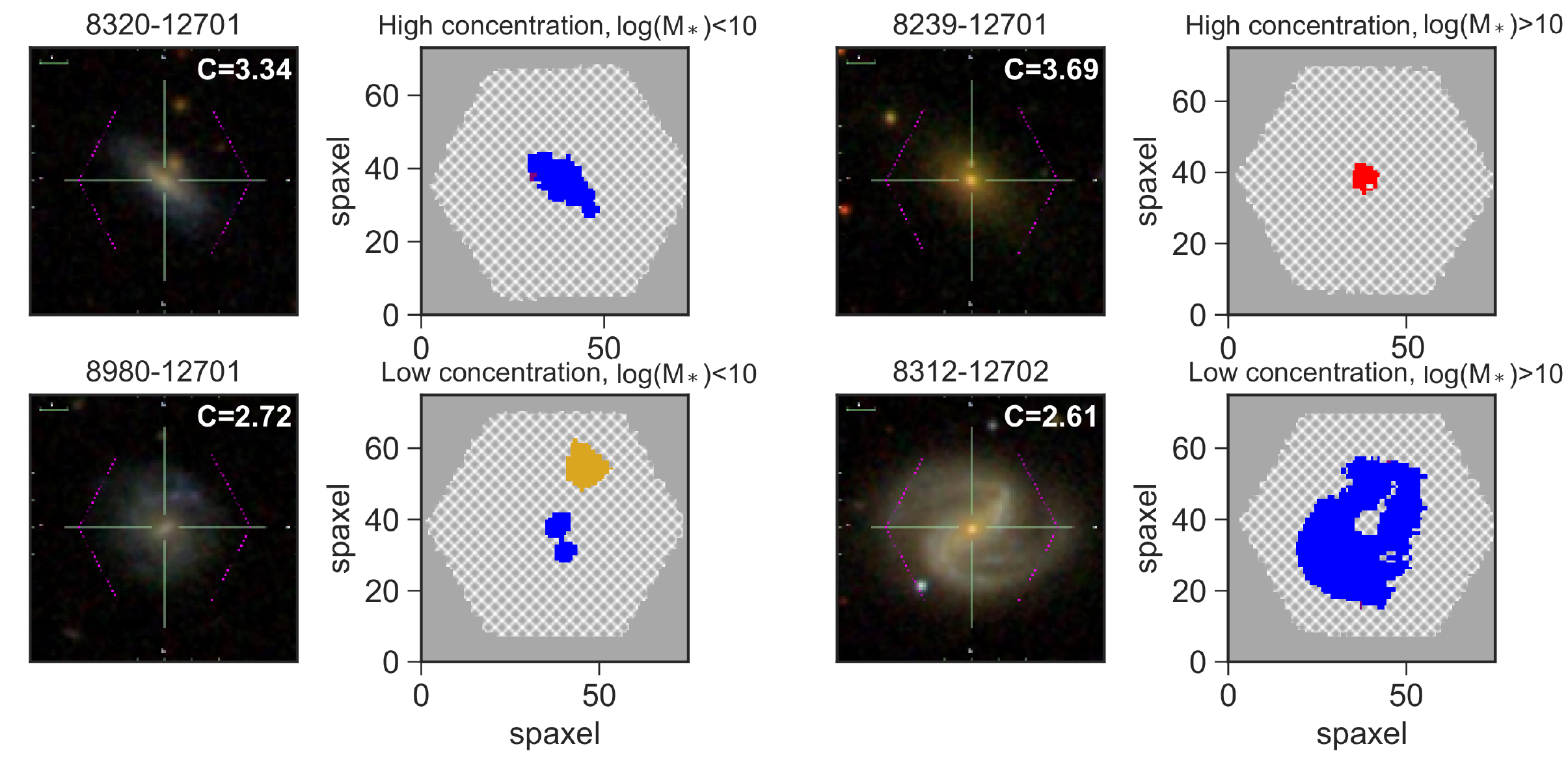}
    \caption{Example galaxies in each of the mass and concentration bins. A low/high value of $A$ refers to low/high concentration. Left: SDSS $gri$ image of an example galaxy showing the MaNGA IFU footprint as the magenta hexagon. Right: a map of the PCA classes in each galaxy. The classes are: quiescent (red), star forming (blue), green valley (green), starburst (yellow) and post-starburst (purple). The grey region indicates no coverage, and the hatched region indicates where spaxels are masked due to low S/N, non-classification due to non-physical values of (PC1, PC2), making in the DAP cubes e.g. due to stellar contamination or because of a high Balmer decrement (H$\alpha$/H$\beta$ $>5.16$).}
    \label{fig:Conc_Mstar_examples}
\end{figure*}

\begin{figure*}
	\includegraphics[width=0.99\textwidth]{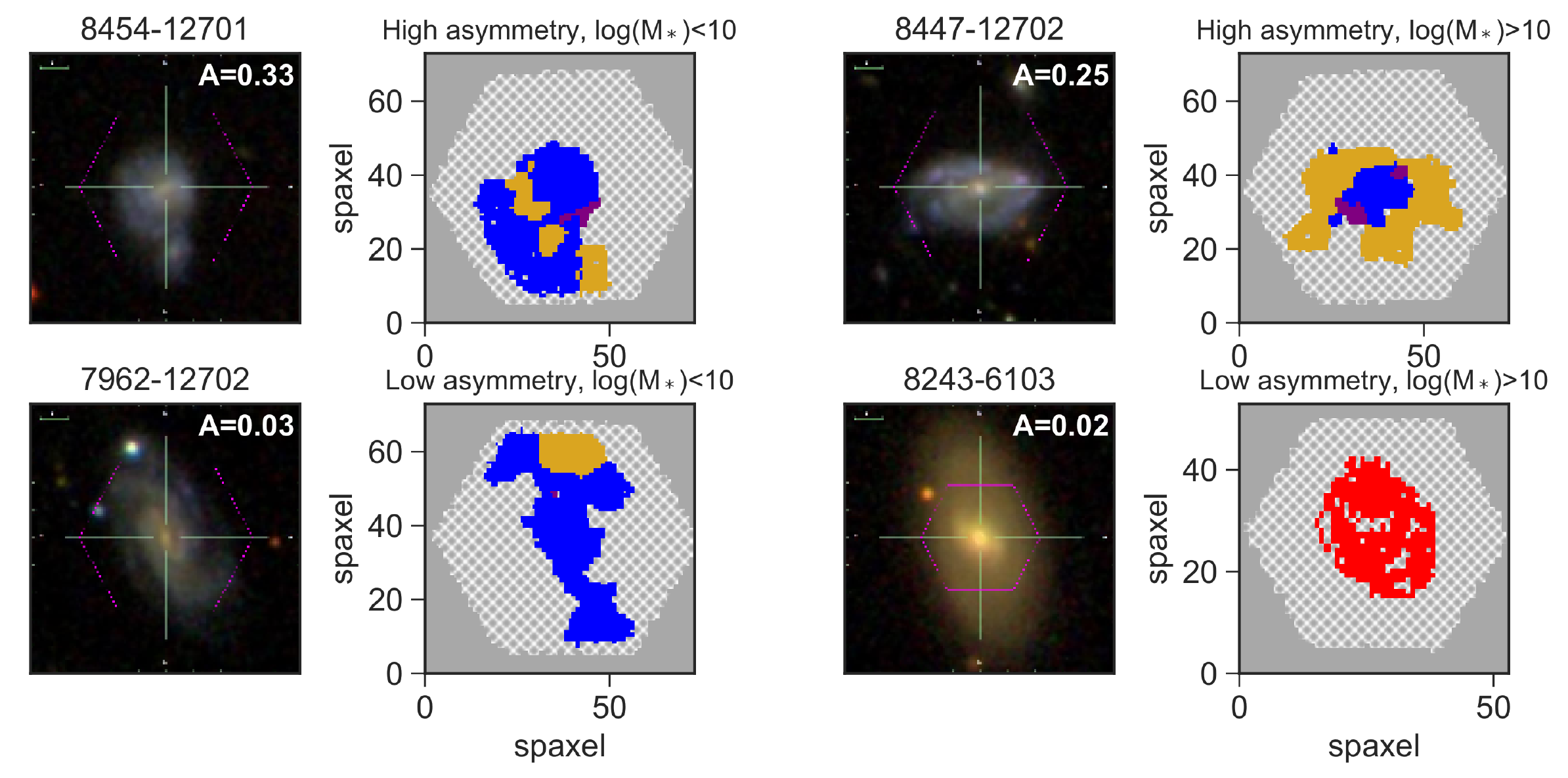}
    \caption{Example galaxies in each of the mass and asymmetry bins. A low/high value of $A$ refers to low/high asymmetry. Left: SDSS $gri$ image of the example galaxy showing the MaNGA IFU footprint as the magenta hexagon. Right: a map of the PCA classes in each galaxy. The classes are: quiescent (red), star forming (blue), green valley (green), starburst (yellow) and post-starburst (purple). The grey region indicates no coverage, and the hatched region indicates where spaxels are masked due to low S/N, non-classification due to non-physical values of (PC1, PC2), making in the DAP cubes e.g. due to stellar contamination or because of a high Balmer decrement (H$\alpha$/H$\beta$ $>5.16$).}
    \label{fig:Asym_Mstar_examples}
\end{figure*}

In Figures~\ref{fig:Conc_Mstar_examples} and \ref{fig:Asym_Mstar_examples} we show examples of galaxies which fall in each concentration, asymmetry and mass bin.

\end{appendix}


\bsp	
\label{lastpage}
\end{document}